\documentclass{JHEP3}

\usepackage{graphicx}

\usepackage{amsmath}
\usepackage{amsfonts}   
\usepackage{amssymb}  
\usepackage{mathrsfs}

\newcommand{\data}{d}
\newcommand{\nuis}{\psi}
\newcommand{\params}{\theta}
\newcommand{\basis}{\Theta}

\newcommand\eg{{\it {e.g.}}}  
\newcommand\etal{{\it {et al.}}}

\newcommand{\be}{\begin{equation}}
\newcommand{\ee}{\end{equation}}
\newcommand{\beq}{\begin{equation}}
\newcommand{\eeq}{\end{equation}}
\newcommand{\bea}{\begin{eqnarray}}
\newcommand{\eea}{\end{eqnarray}}
\newcommand{\gsim}{\lower.7ex\hbox{$\;\stackrel{\textstyle>}{\sim}\;$}}
\newcommand{\lsim}{\lower.7ex\hbox{$\;\stackrel{\textstyle<}{\sim}\;$}}

\newcommand{\mhalf}{m_{1/2}}      \newcommand{\mzero}{m_0}
\newcommand{\tanb}{\tan\beta}
\newcommand{\azero}{A_0}

\newcommand{\cl}{\text{CL}}
\newcommand{\mtpole}{M_t}
\newcommand{\alphas}{\alpha_s(M_Z)^{\overline{MS}}}
\newcommand{\alphaemmz}{\alpha_{\text{em}}(M_Z)^{\overline{MS}}}
\newcommand{\sineff}{\sin^2 \theta_{\rm{eff}}}

\newcommand{\BR}{BR}
\newcommand{\deltaamususy}{\delta a_{\mu}^{\text{SUSY}}}
\newcommand{\brbsgamma}{\BR(\overline{B}\rightarrow X_s\gamma)}
\newcommand{\bsgamma}{b\rightarrow s \gamma }
\newcommand\delmbs{\Delta M_{B_s}}
\newcommand\brbtaunu{\BR(\overline{B}_u\to \tau \nu)}
\newcommand\brbsmumu{\BR(\overline{B}_s\to\mu^+\mu^-)}

\newcommand{\abundchi}{\Omega_\chi h^2}

\newcommand\ps{\mbox{ ps}}
\newcommand{\mev}{\mbox{ MeV}}
\newcommand{\gev}{\mbox{ GeV}}
\newcommand{\tev}{\mbox{ TeV}}

\newcommand{\mhl}{m_h}
\newcommand{\zetah}{\zeta_h}

\newcommand{\amuexpt}{a_{\mu}^{\text{expt}}} 
\newcommand{\amusm}{a_{\mu}^{\text{SM}}}
\newcommand{\gmtwo}{(g-2)_{\mu}}

\newcommand{\ww}{0.49\linewidth}
\newcommand{\qq}{0.24\linewidth}
\newcommand{\Aeff}{A^\text{eff}}

\newcommand{\GA}{\Gamma_\text{ann}}
\newcommand{\Gcap}{\Gamma_\text{capt}}
\newcommand{\fsyst}{f^\text{syst}}
\newcommand{\Ep}{\mathscr{E}}
\newcommand{\Eth}{E^\text{th}}
\newcommand{\Ethmu}{E^\text{th}_\mu}
\newcommand{\bbar}{b\bar{b}}
\newcommand{\tbar}{t\bar{t}}
\newcommand{\rholoc}{\rho_\text{loc}}

\newcommand{\sigsip}{\sigma_p^\text{SI}}
\newcommand{\sigsdp}{\sigma_p^\text{SD}}
\newcommand{\sigsdn}{\sigma_n^\text{SD}}
\newcommand{\Nmu}{N_\mu}
\newcommand{\rpp}[3]{{\it Rept.\ Prog.\ Phys.\ }{\bf #1} (#2) #3}
\newcommand{\iyr}{~\text{yr}^{-1}}
\newcommand{\mchi}{m_\chi}

\title{Prospects for dark matter detection with IceCube in the context of the CMSSM}

\author{Roberto Trotta\\
	Astrophysics Group, Imperial College London \\
	Blackett Laboratory, Prince Consort Road, London SW7 2AZ, UK \\ 
        E-mail: \email{r.trotta@imperial.ac.uk}}

\author{Roberto Ruiz de Austri\\
        Instituto de F\'isica Corpuscular, IFIC-UV/CSIC \\
        Valencia, Spain \\
        E-mail: \email{rruiz@ific.uv.es}}

\author{Carlos P\'erez de los Heros\\
        Department of Physics and Astronomy\\
        Uppsala University\\
        Uppsala, Sweden\\
        E-mail: \email{cph@fysast.uu.se}}

\abstract{ We study in detail the ability of the nominal configuration of the  
IceCube neutrino telescope
(with 80 strings) to probe the parameter space of the Constrained MSSM  
(CMSSM) favoured by
current collider and cosmological data. Adopting conservative  
assumptions about the galactic
halo model and the expected experiment performance, we find that  
IceCube has a probability between
2\% and 12\% of achieving a $5\sigma$ detection of dark matter  
annihilation in the Sun,
depending on the choice of priors for the scalar and gaugino masses and on  
the astrophysical assumptions. We identify
the most important annihilation channels in the CMSSM parameter space  
favoured by current constraints,
and we demonstrate that assuming that the signal is dominated by a  
single annihilation channel can
lead to large systematic errors in the inferred WIMP annihilation  
cross section. We demonstrate that
$\sim 66\%$ of the CMSSM parameter space violates the equilibrium  
condition between capture
and annihilation in the center of the Sun. By cross--correlating our  
predictions with direct
detection methods, we conclude that if IceCube does detect a neutrino  
flux from the Sun at
high significance while direct detection experiments do not find a  
signal above a spin--independent
cross section $\sigsip \gsim 7 \times 10^{-9}$ pb, the CMSSM will be  
strongly disfavoured, given standard astrophysical assumptions for the WIMP distribution. This result is robust with respect to a change of priors. We argue that the proposed low-energy DeepCore extension of IceCube will 
be an ideal instrument to focus on relevant CMSSM areas of parameter space.} 


\preprint{}

\begin{document}

\section{Introduction}\label{sec:intro}

Supersymmetry (SUSY) is the most promising candidate for physics beyond the Standard Model (SM). 
It provides solutions to two of the most challenging questions 
facing particle physics today~\cite{susy-reviews}. Namely, the instability of the Higgs 
mass against radiative corrections (known as the ``fine--tuning problem'') 
and the nature of dark matter. Firstly, the fine--tuning problem is addressed via the cancellation
of quadratic divergences in the radiative corrections to the Higgs
mass. Secondly, the theory predicts the existence of Weakly Interacting Massive Particles (WIMPs). 
Assuming  R-parity conservation, the lightest supersymmetric particle is stable and can have survived 
as a relic from the Big Bang, meeting the necessary requirements to account 
for the the dark matter content of the Universe suggested by observations~\cite{susy-dm-reviews}. \par

Under the above mentioned assumptions, our galaxy could be embedded in a dark halo formed of WIMPs.  
During the last decades a wide range of experiments have been developed with the goal of detecting such WIMPs. 
The overall strategy has followed a double--track approach: on the one hand, ``direct detection experiments'' 
aim at measuring the energy deposited by the interaction of a WIMP from the halo with a nucleus in the target material~\cite{direct-searches-review}. 
On the other hand ``indirect detection experiments'' focus on the detection of the annihilation products of WIMPs, such as cosmic 
rays, neutrinos or $\gamma$ rays, produced by dark matter annihilation from a variety of sources, including the galactic 
center~\cite{Jeltema:2008hf}, satellite dwarf galaxies~\cite{Martinez:2009jh}, and diffuse emission from unresolved sources. 

A promising search technique is to look for an excess of neutrinos from WIMP annihilations in the 
center of heavy bodies, like the Earth or the Sun. The idea behind this possibility is that WIMPs from the halo can be gravitationally 
trapped in orbits in the Solar System and, by scattering off the matter of heavy bodies,  fall below the escape velocity and accumulate 
in their center~\cite{Press:1985ug}. These trapped WIMPs can then annihilate 
with each other to produce quarks, leptons, gauge or Higgs bosons. The decays of some of these particles will produce 
neutrinos, which will escape the body, providing  an excellent experimental signature: they will be directional  
and will have energies from tens to several hundreds GeV depending mainly on the WIMP mass.  In the absence of a signal, the 
existing neutrino telescopes, AMANDA, IceCube, ANTARES and Baikal,  as well as the SuperKamiokande detector have already set 
limits on the neutrino flux from neutralino annihilations in the Sun or Earth~\cite{Achterberg:06a, Ackerman:06a, IceCube:09a, Baikal:96a, SuperK:04a}.

As mentioned above, the supersymmetric framework offers a natural and well motivated candidate for a WIMP, namely the neutralino. The neutralino is the mass eigenstate 
of a mixture of bino, wino (the superpartners of the $B$, $W^0$ gauge bosons) and higgsinos (the superpartners of the $H^0_1$, $H^0_2$ Higgs bosons). Whenever the lightest 
among the four neutralinos is the lightest SUSY particle, it is expected to be stable due to R--parity (in the following, ``neutralino'' always denotes the lightest 
of the 4 neutralino eigenstates). Therefore the neutralino provides a stable, neutral, massive (typical masses range from tens of GeV up to several TeV) particle that 
can account for the  cold dark matter required to fit cosmological observations. 

Despite its potentiality, low-energy SUSY has a fundamental question yet to be answered; namely, how it is broken. 
Along the past years, several SUSY--breaking mechanisms have been proposed in the literature. 
Perhaps the most appealing one regards to gravity as the mean of communication among a hidden sector 
where SUSY breaking occurs and the visible sector \cite{Nilles:1983ge}. This scenario is commonly called Supergravity, 
and a simple realization of it is the so called Constrained Minimal Supersymmetric Standard Model (CMSSM) \cite{kkrw94,sugra-reviews}. 
It is defined in terms of five free parameters: common scalar ($\mzero$), gaugino ($\mhalf$) and
tri--linear ($\azero$) mass parameters (all specified at the GUT scale) plus the ratio of Higgs vacuum expectation values $\tanb$
and $\text{sign}(\mu)$, where $\mu$ is the Higgs/higgsino mass parameter whose square is computed from the conditions of
radiative electroweak symmetry breaking.

There is a vast literature devoted to the study of the phenomenology of the CMSSM, based on current collider results and 
data from dark matter searches. The usual procedure to explore the model's parameter space has been to evaluate the likelihood of the observables on a a fixed grid, often encompassing only 2 or 3 dimensions at the time~\cite{grid-cmssm}. 
This method has several limitations. Firstly, the number of points required scales as $k^N$, where $N$ is the number 
of the model's parameters and $k$ the number of points for each of them. Therefore this technique becomes highly 
inefficient for exploring with sufficient resolution parameter spaces of even modest dimensionality. Secondly, some features of the parameter space can easily be missed by not setting a fine enough resolution. 
Thirdly, extra sources of uncertainties (\eg, those due to the lack of precise knowledge of SM parameter values) 
and relevant external information (\eg, about the parameter range) are difficult to accommodate.

More recently a new approach based on Bayesian methods (for a recent review of Bayesian methods, see, e.g. \cite{Trotta:2008qt}) has been successfully applied 
to the exploration of supersymmetry phenomenology~\cite{bg04,al06,rtr1,alw06,rrt2,rrt3,Trotta:2006ew,rrts,Allanach:2008tu,BenModelComp,Allanach:2008iq,tfhrr:2008,nuhm1,fhrrt2,AbdusSalam:2009qd}. Bayesian methods coupled with Markov Chain Monte Carlo (MCMC) technology are superior in many
respects to traditional grid scans of the parameter space.  Firstly, they are much more efficient, in that the computational effort required to explore a
parameter space of dimension $N$ scales roughly proportionally with
$N$ for standard MCMC methods. The more recent ``nested sampling'' algorithm~\cite{SkillingNS, Feroz:2007kg}
 offers an efficiency which is roughly constant with $N$, thus opening up the possibility of exploring parameter spaces of high dimensionality such as the general MSSM. Secondly, the Bayesian approach allows one to incorporate easily  into the final inference all relevant sources of
uncertainty. For a given SUSY model one can include relevant SM
(nuisance) parameters and their associated experimental errors, with
the uncertainties automatically propagated to give the final
uncertainty on the SUSY parameters of interest.  Thirdly, another key advantage is the possibility
to marginalize (i.e., integrate over) additional (``hidden'')
dimensions in the parameter space of interest with essentially no
computational effort. Finally, a key feature is the ability to obtain probability distributions for any function of the input parameters, which makes it possible to generate quantitative statistical predictions within a given SUSY model for any observable quantity of interest. This most useful feature is exploited in this paper to produce statistical predictions for the capability of IceCube to constrain the CMSSM. On the other hand, the priors issue, considered as a caveat of the Bayesian approach, has now been thoroughly addressed in this context~\cite{tfhrr:2008,bclw07,allanach06,ccr08}.

Previous works have considered the detection prospects for neutrino telescopes~(e.g., \cite{Kamionkowski:1994dp,Bergstrom:1998xh,Wikstrom:2009kw}), but without adopting a statistical framework allowing to quantify the probability of discovery within a given theoretical model. More recently, Refs.~\cite{Allanach:2008iq,Bruch:2009rp} have employed a similar technique as the one proposed here, but presented their predictions in terms of muon flux at the detector location. 
There is a vast literature concerning indirect detection signatures of neutrino from the Sun, including models beyond the CMSSM, see e.g.~\cite{NuStudies} (we refer to \cite{direct-searches-review} and references therein for a more exhaustive list of works on the subject).   In this study we improve on previous works by focusing our analysis on the number of muon events in the detector, and by showing that this procedure is immune to potentially large systematic errors that can affect the comparison between theory and data when the dominant annihilation channel is unknown. 

%

The paper is organized as follows. In section~\ref{sec:method} we 
introduce our scanning methodology and our statistical approach, including the data used to constrain the CMSSM parameter space. We then review the basic elements about indirect WIMP searches using neutrino telescopes (section~\ref{sec:muflux}). We present our findings about indirect detection prospects for IceCube and analyze in detail the contribution to the expected signal from several annihilation channels (section~\ref{sec:IDresults}). We compare the indirect detection prospects with direct detection methods in section~\ref{sec:DD} and we discuss the dependency of our results on priors assumptions in section~\ref{sec:prior_change}. We present our conclusions in section~\ref{sec:conclusions}.


\section{Methodology} \label{sec:method}

\subsection{Statistical framework and scanning method}

We denote the parameter set of the CMSSM introduced above, ($\mzero$, $\mhalf$, $\azero$ and $\tanb$), by $\params$ (we fix sgn($\mu$) 
to be positive, motivated by arguments of consistency with the measured anomalous magnetic moment of the muon), while $\nuis$ 
denotes the relevant SM quantities that enter in the calculation of the observable quantities (so--called ``nuisance parameters''), namely
\begin{equation}
\nuis \equiv \{ \mtpole, m_b(m_b)^{\overline{MS}}, \alphas , \alphaemmz \} \, ,
\end{equation}
where $\mtpole$ is the pole top quark mass, $m_b(m_b)^{\overline{MS}}$ is the 
bottom quark mass at $m_b$, while $\alphaemmz$ and $\alphas$ are the 
electromagnetic and the strong coupling constants at the $Z$ pole mass $M_Z$, 
the last three evaluated in the $\overline{MS}$ scheme.
We denote the full 8--dimensional set of parameters by 
\be
\basis = (\params,\nuis).
\label{basis:eq}
\ee
The cornerstone of Bayesian inference is  Bayes' Theorem, which reads
\be \label{eq:bayes}
 p(\basis | \data) = \frac{p(\data |
\basis) p(\basis)}{p(\data)}. \ee
The quantity $p(\basis | \data)$ on the l.h.s. of eq.~\eqref{eq:bayes}
is called the {\em posterior}, while on the r.h.s., the quantity $p(\data |
\basis)$ is the likelihood (when taken as a function $\basis$ for fixed data, $d$). In this paper, we use the same 
likelihood function as given in Sec.~3.1 of ref.~\cite{rtr1}, with updated values for the experimental constraints as applicable (see Table~\ref{tab:obs}).  The quantity $p(\basis)$ is the {\em prior} which 
encodes our state of knowledge about the values of the parameters $\basis$ before we see the data. The state of knowledge 
is then updated to the posterior via the likelihood.  Finally, the quantity in the denominator is called
{\em evidence} or {\em model likelihood}. If one is interested in constraining the model's parameters, the evidence is merely a
normalization constant, independent of $\basis$, and can therefore be dropped.

In order to explore efficiently the posterior of Eq.~\eqref{eq:bayes}, we adopt the MultiNest~\cite{Feroz:2007kg} algorithm as 
implemented in the \texttt{SuperBayeS} code~\cite{tfhrr:2008}. MultiNest provides an extremely efficient sampler even for likelihood functions defined over a parameter space of large dimensionality with a 
very complex structure. This aspect is very important for the CMSSM, as previous MCMC scans have revealed that the 8-dimensional likelihood 
surface is very fragmented and that it presents many finely tuned regions that are difficult to explore with conventional MCMC (and almost 
impossible to find with conventional grid scans). By explicitly including the SM parameters and then integrating over them at the end, our scans automatically account for the full uncertainty associated with the current experimental errors on those quantities. In particular it has been shown that it is important to include the impact of the uncertainty in the top mass to compute accurately the CDM relic abundance~(see Figure 3 in~\cite{rrt3}).

\subsection{Choice of priors} 

In order to perform the scan over the CMSSM and SM parameters, we need to specify a prior for the l.h.s~of Bayes' theorem, Eq.~\eqref{eq:bayes}. The role of the prior is to define a statistical measure on the parameter space. In principle, when the likelihood is strongly constraining (i.e., for accurate data) the posterior is dominated by the likelihood and the prior choice is irrelevant, as the information in the likelihood completely overrides the information in the prior (for an illustration, see Fig.~2 in~\cite{Trotta:2008qt}). However, it has been shown that this is presently not the case for the CMSSM, i.e. different (plausible) choices of priors do lead to different posteriors and hence different inferences on the parameter space (for a detailed analysis, see~\cite{tfhrr:2008}). The resolution of this issue will come from future data, and in particular LHC measurements of the SUSY mass spectrum, which will conclusively resolve the ambiguity brought about by the choice of priors~\cite{LHC:inprep}. 
Whenever several prior choices appear plausible based on physical considerations, it is important to verify to which extent one's statistical conclusions depend upon the choice of priors. 
In this work, we present our main results for the choice of so--called ``log priors'', namely priors that are flat in $\log\mhalf$ and $\log\mzero$, and flat in $\tanb$ and $A_0$. The prior ranges are chosen as to cover the region where the likelihood is significantly non--zero, and span the following intervals\footnote{We also follow the treatment of~\cite{tfhrr:2008} for the priors on the SM parameters, although this choice does not represent an issue as those quantities are directly constrained by experiment and thus do not suffer from prior--ambiguity.}:
\begin{align}
& 1.69 \leq \log\frac{\mzero}{1 \gev},\log\frac{\mhalf}{1 \gev} \leq 3.60 \\
& (\text{corresponding to } 50 \mev \leq \mzero, \mhalf \leq 4\tev) \\
& -7 \tev \leq A_0 \leq 7 \tev \\
& 2 \leq \tanb \leq 62.
\end{align} 

Our choice of log priors is dictated by both physical and statistical considerations. From the physical point of view, adopting a prior that is flat in the log of the masses means assuming {\em a priori} that all orders of magnitudes have the same probability, e.g. that the probability of $10 \gev < \mzero < 100 \gev$ is the same as the probability  $100 \gev < \mzero < 1 \tev$. This is appealing because it does not single out any order of magnitude for the mass scale in the problem. From the statistical point of view, it was shown in~\cite{tfhrr:2008} that this choice of priors produces a posterior probability that is close to what one would get from a naive grid--scan of the likelihood, and maximising the likelihood along the hidden dimensions (so--called ``profile likelihood''). A detailed statistical analysis shows that the average quality of fit of a log prior scan is much better than the average quality of fit assuming flat priors~\cite{tfhrr:2008}, i.e. priors flat in $\mzero, \mhalf$. For all those reason, we adopt the log prior choice as our default prior for the main results presented in the following. We comment on the sensitivity of our results to changes in the prior choice in section~\ref{sec:prior_change}.

\subsection{Observables and data}

The experimental values of collider and cosmological observables that we employ to constrain the CMSSM parameter space are listed in
Table~\ref{tab:obs}, while the constraints on the SM nuisance parameters are given in Table~\ref{tab:nuisance}. We refer 
to~\cite{rtr1,rrt2,tfhrr:2008} for details about the computation of each quantity and for justification of the theoretical errors adopted, 
as well for a detailed description of the likelihood function. In particular, we include the measurement of the anomalous magnetic moment 
of the muon based on $e^+e^-$ data, which gives a $3.2\sigma$ discrepancy with the SM predicted value~\cite{gm2}. 
As regards $\brbsgamma$ (which has been recently shown to provide an important constraint~\cite{rrt2}), for the new SM prediction
we employ the value of $(3.12\pm 0.21)\times10^{-4}$.  We compute the SUSY contribution to $\brbsgamma$ following the procedure outlined in
Refs.~\cite{dgg00,gm01} which was extended in Refs.~\cite{or1+2,for1+2} to the case of general flavor mixing.  

Regarding cosmological constraints, we use the determination of the relic abundance of cold dark matter based on the 5-year data from
WMAP~\cite{wmap5yr} to constrain the relic abundance $\abundchi$ of the lightest neutralino (assumed to be the sole constituent of dark matter).  
Also, points that do not fulfil the conditions of radiative electroweak symmetry breaking and/or give non-physical (tachyonic) solutions are discarded.

\begin{table}[t]
\centering
\begin{tabular}{|l | l l l | l|}
\hline
Observable &   Mean value & \multicolumn{2}{c|}{Uncertainties} & ref. \\
 &   $\mu$      & ${\sigma}$ (exper.)  & $\tau$ (theor.) & \\\hline
 $M_W$     &  $80.398\gev$   & $25\mev$ & $15\mev$ & \cite{lepwwg} \\
$\sineff{}$    &  $0.23153$      & $16\times10^{-5}$
                & $15\times10^{-5}$ &  \cite{lepwwg}  \\
$\deltaamususy \times 10^{10}$       &  29.5 & 8.8 &  1.0 & \cite{gm2}\\
 $\brbsgamma \times 10^{4}$ &
 3.55 & 0.26 & 0.21 & \cite{hfag} \\
$\delmbs$     &  $17.77\ps^{-1}$  & $0.12\ps^{-1}$  & $2.4\ps^{-1}$
& \cite{cdf-deltambs} \\
$\brbtaunu \times 10^{4}$ &  $1.32$  & $0.49$  & $0.38$
& \cite{hfag} \\
$\abundchi$ &  0.1099 & 0.0062 & $0.1\,\abundchi$& \cite{wmap5yr} \\\hline\hline
   &  Limit (95\%~\cl)  & \multicolumn{2}{r|}{$\tau$ (theor.)} & ref. \\ \hline
$\brbsmumu$ &  $ <5.8\times 10^{-8}$
& \multicolumn{2}{r|}{14\%}  & \cite{cdf-bsmumu}\\
$\mhl$  & $>114.4\gev$\ (for an SM-like Higgs)  & \multicolumn{2}{r|}{$3 \gev$}
& \cite{lhwg} \\
$\zetah^2$
& $f(m_h)$\ (see~\cite{rtr1})  & \multicolumn{2}{r|}{negligible}  & \cite{lhwg} \\
$m_{\tilde{q}}$ & $>375$ GeV  & & 5\% & \cite{pdg07}\\
$m_{\tilde{g}}$ & $>289$ GeV  & & 5\% & \cite{pdg07}\\
other sparticle masses  &  \multicolumn{3}{c|}{As in Table~4 of
  Ref.~\cite{rtr1}.}  & \\ \hline 
\end{tabular}
\caption{Summary of the observables used in the analysis to constrain the CMSSM parameter space. Upper part:
Observables for which a positive measurement has been
made. $\deltaamususy=\amuexpt-\amusm$ denotes the discrepancy between
the experimental value and the SM prediction of the anomalous magnetic
moment of the muon $\gmtwo$. 
Lower part: Observables for which only limits currently
exist.  
\label{tab:obs}}
\end{table}

\begin{table}[t]
\centering    .

\begin{tabular}{|l | l l | l|}
\hline
SM (nuisance)  &   Mean value  & \multicolumn{1}{c|}{Uncertainty} & Ref. \\
 parameter &   $\mu$      & ${\sigma}$ (exper.)  &  \\ \hline
$\mtpole$           &  172.6 GeV    & 1.4 GeV&  \cite{topmass:mar08} \\
$m_b (m_b)^{\overline{MS}}$ &4.20 GeV  & 0.07 GeV &  \cite{pdg07} \\
$\alphas$       &   0.1176   & 0.002 &  \cite{pdg07}\\
$1/\alphaemmz$  & 127.955 & 0.03 &  \cite{Hagiwara:2006jt} \\ \hline
\end{tabular}
\caption{Experimental mean $\mu$ and standard deviation $\sigma$ 
 adopted for the likelihood function for SM (nuisance) parameters,
 assumed to be described by a Gaussian distribution.
\label{tab:nuisance}}
\end{table}


\section{Indirect WIMP searches with neutrino telescopes} 
\label{sec:muflux}

 As mentioned in the introduction, indirect searches for dark matter are based on the search for a neutrino signal 
from the center of the Earth or the Sun over the known atmospheric neutrino background. 
The most sensitive searches to date are those carried out by neutrino telescopes. These are detectors that use huge 
instrumented volumes of natural ice or water both as target and Cherenkov medium to detect the muons\footnote{Electron or tau-induced 
 particle cascades, in the case of incoming $\nu_e$ or $\nu_{\tau}$, do not provide a good enough angular resolution for the 
studies carried out in this paper, so we concentrate here on the $\nu_{\mu}$ channel only.} produced in neutrino-nucleon interactions in or around the detector. 
An array of photomultiplier tubes detects the Cherenkov photons with nanosecond resolution and allows for the reconstruction 
of the muon trajectory, and hence of the incoming neutrino direction. The angular resolution depends on the incoming neutrino energy, and it is typically a degree at TeV energies. In this paper we investigate the signal coming from neutralino annihilations in the center of the Sun.

\subsection{Neutrinos from WIMPs annihilation in the Sun}

From an experimental point of view, the relevant observable is the total number of neutrino--induced muons from the direction of the Sun per unit time, N$_\mu$, with 
an energy above a given threshold $\Ethmu$. This N$_\mu$  is given by the convolution of the muon effective area of the detector, $\Aeff_\mu(E_\mu)$, with the neutrino--induced differential muon flux at its location, $\frac{d \Phi_\mu}{d E_\mu}$:
\be \label{eq:muevent}
N_\mu =  \int^{+\infty}_{\Eth_\mu} d E_\mu \Aeff_\mu(E_\mu) 
\frac{d \Phi_\mu}{d E_\mu}.
\ee
 The differential flux of neutrino--induced muons reaching the detector is computed from the differential neutrino flux by accounting for both the production cross section of muons from neutrinos and their propagation in ice.  The differential parent neutrino flux, $\frac{d \Phi_\nu}{d E_\nu}$, is given by 
\be \label{eq:nudflux}
\frac{d \Phi_\nu}{d E_\nu} = \frac{\GA}{4 \pi D^2} \sum_i B^i
\frac{d N^i_\nu}{d E_\nu}, 
\ee
where $\GA$ is the annihilation rate, $D$ is the distance from the source to the detector, the sum runs over all WIMP pair annihilation 
final states and  $B^i$ are the branching ratios for each channel, each giving rise to the differential neutrino spectrum $\frac{d N^i_\nu}{d E_\nu}$. There is a total of 29 channels potentially contributing to the sum in Eq.~\eqref{eq:nudflux}, although the most important ones in the context of the CMSSM (as we show below) involve the final states $W^+W^-$, $ZZ$, $\tau^+\tau^-$,  $\bbar$ and, when kinematically allowed, $\tbar$. The energy range of neutrinos produced by WIMP annihilations is bounded by the maximum WIMP mass, which in the case of the CMSSM is of order $\lsim 1$ TeV. 

WIMPs from the galactic halo accrete in the center of the Sun by gravitational capture and their 
number is depleted by annihilation. The capture rate $\Gcap \propto \sigma_s \rholoc$ is proportional to the scattering cross section $\sigma_s$ between a WIMP and a nucleus and to the local WIMP density, $\rholoc$. In order to compute it, one needs to specify both a particle physics model (determining $\sigma_s$) and a halo model, giving the local halo density, $\rholoc$, and the velocity distribution of WIMPs. In this paper we assume an NFW profile with $\rholoc = 0.3$ GeV/cm$^3$ and a Maxwell--Boltzmann velocity distribution with mean $\bar{v} = 270$ km/s. Assuming a constant capture rate, one can compute the time--dependent density of WIMPs in the Sun and from there their annihilation rate, $\GA$, which is given by 
\be \label{eq:GA}
\GA = \frac{\Gcap}{2} \tanh^2 \left ( \frac{t}{\tau} \right ),
\ee
where $\tau = 1/\sqrt{\Gcap C_a}$, where $C_a \propto \langle \sigma_a v \rangle$ is proportional to the velocity--averaged annihilation cross section. 
For timescales $t \gg \tau$, equilibrium between annihilation and capture is established, and thus $\GA \approx \Gcap/2$. This is an important issue for indirect searches since only if capture and 
annihilation are in equilibrium is the neutrino flux from neutralino annihilations at ``full strength''. Conversely, if equilibrium is not reached, $\GA \ll \Gcap$ and the neutrino signal is correspondingly weaker. In our approach, we compute $\Gcap, C_a$ from our theoretical model and galactic halo model, and $\GA$ from Eq.~\eqref{eq:GA}, assuming $t = 4.76$ Gyr for the age of the solar system. Thus the equilibrium condition can be explicitly checked, see section~\ref{sec:equilibrium}.

 So far, none of the neutrino telescopes in operation has detected any excess of neutrinos from the Sun. This lack of signal has allowed to set limits on the muon 
flux from neutralino annihilations in the Sun, as well as on the annihilation rate of neutralinos. The most stringent limits to date are those from 
the 22-string configuration of IceCube obtained with 104 days of livetime during 2007~\cite{IceCube:09a}. In the light of this situation, here we focus on the expected sensitivity to muons of the final IceCube configuration to describe the potential of km$^3$ neutrino telescopes to probe the CMSSM parameter space.

\subsection{The IceCube neutrino telescope}
\label{sec:icecube}
 In this work we will use the proposed original geometry of the IceCube neutrino telescope~\cite{IceCube:04a} to exemplify the impact that kilometer-scale neutrino detectors 
can make in constraining the CMSSM through  searches for an excess in the neutrino flux from the Sun's direction. 
 The baseline IceCube geometry consists of 80 strings with 60 digital optical modules each, deployed at 
depths between 1450~m - 2450~m near the geographic South Pole. The optical modules are vertically separated by 17 m, while the strings are arranged in a triangular 
grid with an inter-string separation of 125 m. The instrumented volume of ice covers 1 km$^3$.  Each module contains a 25-cm Hamamatsu photomultiplier tube with 
electronics for in-situ digitization and timing of the photomultiplier signals~\cite{IceCube:09b}. \par
\begin{figure}[t]
\begin{center}
\includegraphics[width=\ww]{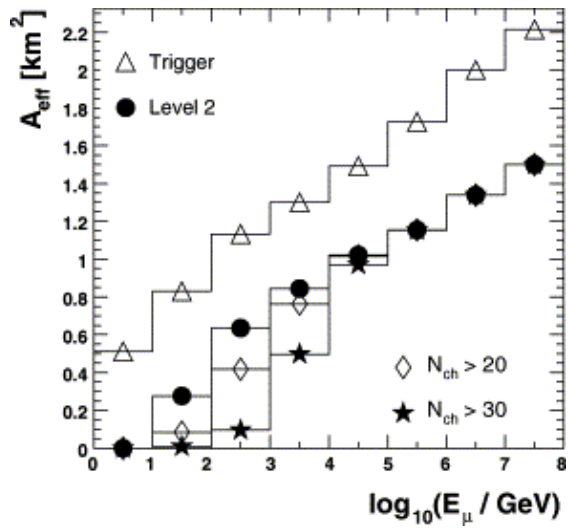}
\caption[test]{Effective area of IceCube (final configuration with 80 strings, from~\cite{IceCube:04a}) as a function of muon energy. Triangles denote trigger level, while black dots (adopted in this work) are the effective area after a simple event selection 
based on track quality parameters. Diamonds and stars represent further energy cuts. See~\cite{IceCube:04a} for details.}
\label{fig:A_eff}
\end{center}
\end{figure}

 IceCube is designed to detect the Cherenkov radiation of secondaries produced in neutrino interactions in the vicinity of the detector. 
Although its geometry has been optimized to detect ultra-high energy ($>$TeV) neutrinos from potential cosmic sources with 
sub-degree angular resolution, its energy and angular response are still adequate at ${\mathcal{O}}(100)$ GeV to perform searches 
for dark matter, where the typical neutrino energies are of that order or less.
 A measure of the efficiency of a neutrino telescope to detect muons is given by the effective detector area, defined as 
\be \label{eq:A_eff}
\Aeff_\mu(E_{\mu})\,=\,\frac{N_\text{sel}}{N_\text{gen}}\times A^\text{gen}(E_{\mu}),
\ee
where $N_\text{sel}$ is the number of events that pass the analysis selection cuts from a test sample of $N_\text{gen}$ generated muons 
that cross a fiducial area  $A^\text{gen}$. The effective area depends on the muon energy, $E_{\mu}$, and the incident angle.
For the studies carried out in this paper we have taken the zenith angle averaged effective area of IceCube as a function of 
muon energy after some simple track quality selection criteria, as published in~\cite{IceCube:04a} and shown here in Fig.~\ref{fig:A_eff} for completeness (black dots). \par

Recently, an addition of six strings forming a denser core in the middle of the IceCube array has been proposed in order to lower the 
neutrino energy threshold of the detector to about  10~GeV~\cite{IceCube:09c}, the so--called ``DeepCore'' configuration. This change 
of the originally proposed geometry will significantly increase the sensitivity of IceCube to searches for low mass WIMPs. 
Since performance studies of DeepCore are still under way at the time of this writing, we have taken a conservative approach and 
use the 80-string geometry of IceCube as the benchmark for our calculations. \par

 In order to make statistically sound statements about the capability of IceCube to detect or rule out parts of the CMSSM, 
 the expected signal obtained through Eq.~\eqref{eq:muevent} has to be compared with the background consisting, in the ideal detector 
case, of atmospheric neutrino--induced events and of an irreducible background of solar atmospheric neutrinos. We neglect here any 
contamination from misreconstructed atmospheric muon tracks, assuming the IceCube analysis can reject such tracks during the analysis 
steps. The signal and the background have a different energy and angular dependency, which can be exploited in the data analysis to 
distinguish between them. The signal from WIMP annihilations in the Sun is expected to come from a cone of angular size $\Delta\Psi$  
from the direction of the Sun, while the atmospheric neutrino background is expected to be, to first approximation, isotropic. 
In this paper we do not employ the signal's energy dependency, but we restrict our analysis to a simple counting scenario, assuming 
that observations of atmospheric neutrinos coming from directions away from the Sun can be used to estimate the number of background 
events $N_b$ above the threshold energy in the signal region. 
 For our estimation of the background we follow the results of the simple analysis presented in~\cite{IceCube:04a}, 
according to which $\sim 10^5$ atmospheric neutrinos are expected per year from the whole hemisphere. We will here make another 
conservative choice and take for our our analysis a cone of $\Delta\Psi = 3.0 ^\circ$ around the Sun position. IceCube is expected to 
have a better angular resolution than that, especially for tracks of energy ${\mathcal{O}}(100)$ GeV and above, so a dedicated WIMP search  
analysis with the complete detector is likely to achieve a better S/N than what we use here. Rescaled to 1 yr (365 d) of detector live time, 
we estimate about 34 events per year from atmospheric neutrinos from the direction of the Sun. An irreducible high-energy neutrino 
background from the Sun are the neutrinos produced by cosmic ray interactions in the corona. Estimations of such effect show that 
the expected number of events in a km$^3$ detector are a few per year~\cite{Ingelman:96a, Fogli:2006jk}. We have used figure~8 in~\cite{Fogli:2006jk} to get an estimation of the number of such events that we would expect in IceCube. Since the authors assume a flat 
effective area of 1 km$^2$, independent of energy, we have rescaled the prediction one can extract from their 
figure~8 for an energy threshold of 50 GeV, to a more realistic effective area of 0.2 km$^2$ at those energies, based on 
figure~\ref{fig:A_eff} in this paper. This leads to estimating $~ 6$ events from the corona per year. Since both the prediction of the 
number of atmospheric neutrinos, as well as the prediction of the number of corona events are affected by the uncertainty in the 
normalization of the cosmic ray spectrum, we adopt here a total number of background events from within  3$^\circ$ from the Sun direction 
in a year of data taking of $N_b = 40\iyr$, and we will not specify any further whether they are corona or atmospheric events. We neglect in the rest of this paper the  uncertainty in the estimation of $N_b$ and we thus assume that its value is perfectly known. We will improve on these approximations in a future work. 

We then estimate the signal--to--noise of a WIMP--induced muon rate $N_\mu$ over the background as 
 \be \label{eq:SN}
 \text{S/N} = \frac{\Nmu}{\sqrt{N_b}}.
 \ee
The parameter space region accessible to IceCube is defined by the criterion S/N$>5$, i.e., as the region where IceCube would obtain an approximate $5\sigma$ detection of the WIMP signal. This corresponds to an event rate due to WIMPs annihilation $\Nmu > 31.6\iyr$ On the other hand, we can also consider the case where a measurement compatible with background-only events is obtained. For definiteness, we assume that the observed number of events is exactly equal to the predicted background rate, $N_b$. We use the Feldman and Cousins construction~\cite{Feldman:1997qc} to derive the 90\% upper exclusion limit for the signal event rate, $N_{90}$, and we find $N_{90} = 11.5\iyr$. In other words, in absence of a signal the parameter space region predicting $\Nmu > N_{90}$ would be excluded at the 90\% CL. When comparing the CMSSM predictions with the IceCube reach, we use the S/N$>5$ criterium as discovery threshold and the $\Nmu > N_{90}$ criterium as the projected exclusion limit. Since the CMSSM predicted event rates stretch over more than 10 orders of magnitude (with the majority of the parameter space below either the detection or exclusion thresholds), effectively it makes almost no difference which one of the two thresholds one adopts to define the reach of the detector. For definiteness, when we discuss ``the reach'' of IceCube we always refer to the region delimited by the discovery criterium S/N$>5$.

\section{Indirect detection prospects with IceCube in the CMSSM framework}
\label{sec:IDresults}

We have now assembled all the relevant elements to predict the number of muon events in IceCube from WIMPs annihilation 
in the Sun. Our procedure is as follows. We have modified the publicly available \texttt{SuperBayeS} code~\cite{superbayes} to link it with DarkSusy~5.0.4~\cite{darksusy}. 
For every point in the CMSSM parameter space visited by our sampling algorithm, DarkSusy~\cite{darksusy}
computes the annihilation and capture rates, $\GA$, $\Gcap$, and the branching ratios into the different annihilation channels, $B^i$ 
in Eq.~\eqref{eq:nudflux}. These are then used to obtain the number of neutrino--induced events at the location of the detector by 
interpolating a series of tables with the yields for each annihilation channel obtained via the WimpSim 
code~\cite{wimpsim,Blennow:2007tw}. WimpSim employs a Monte Carlo simulation to generate the three-flavour neutrino flux at the 
center of the Sun from WIMP annihilations and to propagate it to the Earth including full-flavour oscillations. It has been shown 
that including the effect of neutrino oscillations can have a very substantial impact on the predicted neutrino flux at the Earth 
for some channels~\cite{Blennow:2007tw,Cirelli:2005gh}. Here we assume a  normal neutrino mass hierarchy 
and the neutrino mixing angle $\theta_{13} = 0$. Residual uncertainties in the neutrino physics (in particular, an inverted mass 
hierarchy and/or a non--zero $\theta_{13}$ mixing angle~\cite{Fogli:2008jx}) might lead to modifications of some of the 
yields~\cite{Blennow:2007tw}, although a recent study has not found a significant impact~\cite{Wikstrom:2009kw}. We adopt a muon 
effective area, $\Aeff_\mu$,  obtained by interpolating the values of $\Aeff_\mu(E_\mu)$  
in Fig.~\ref{fig:A_eff} in the energy range $\Ethmu = 50 \gev < E < 1 \tev$, and we use it to perform the convolution with the predicted 
muon spectrum at every point visited in the CMSSM parameter space, according to Eq.~\eqref{eq:muevent}.

Our scanning procedure produces a set of samples $\{ \basis^{(1)}, \basis^{(2)}, \dots \}$ in the CMSSM parameter space whose density is proportional to their posterior 
probability given current data, Eq.~\eqref{eq:bayes}. For any derived quantity $f(\basis)$ that is a function of our CMSSM parameters (such as the muon flux, or the 
annihilation and capture rates, computed as described above) we can simply plot the distribution of $f(\basis^{(i)})$ ($i=1,\dots, N$) and the latter will automatically 
inherit the statistical properties of the samples, i.e. the distribution of derived quantities can also be interpreted in a statistical way. This means that we can make 
probabilistic statements about all sorts of derived quantities (and their correlations) that are a function of our CMSSM parameters and/or of any of the derived observables 
(fluxes, rates, etc).

\subsection{Expected number of events and probability of discovery}

\begin{figure}[t!]
\begin{center}
\includegraphics[width=\ww]{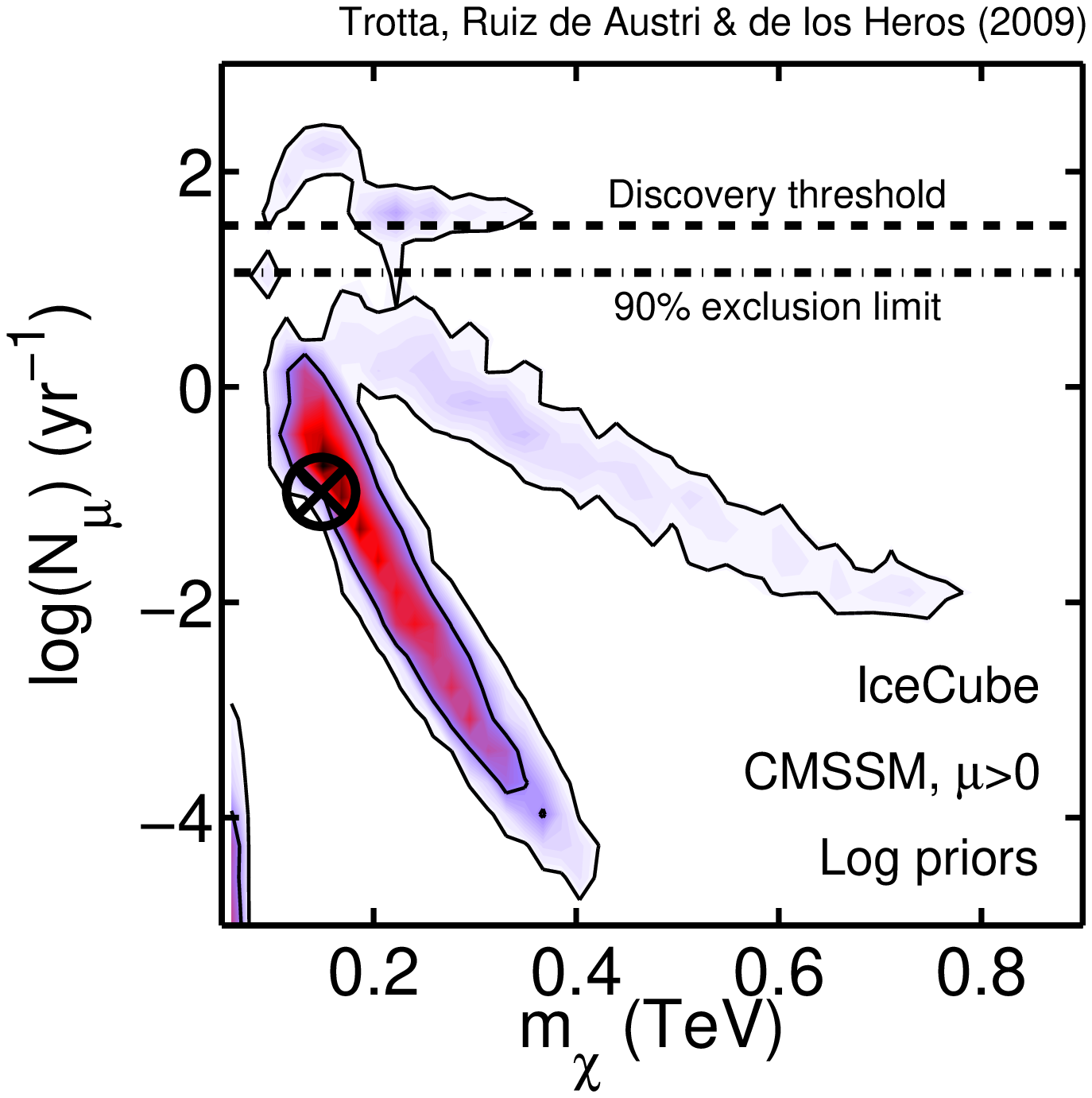}
\includegraphics[width=\ww]{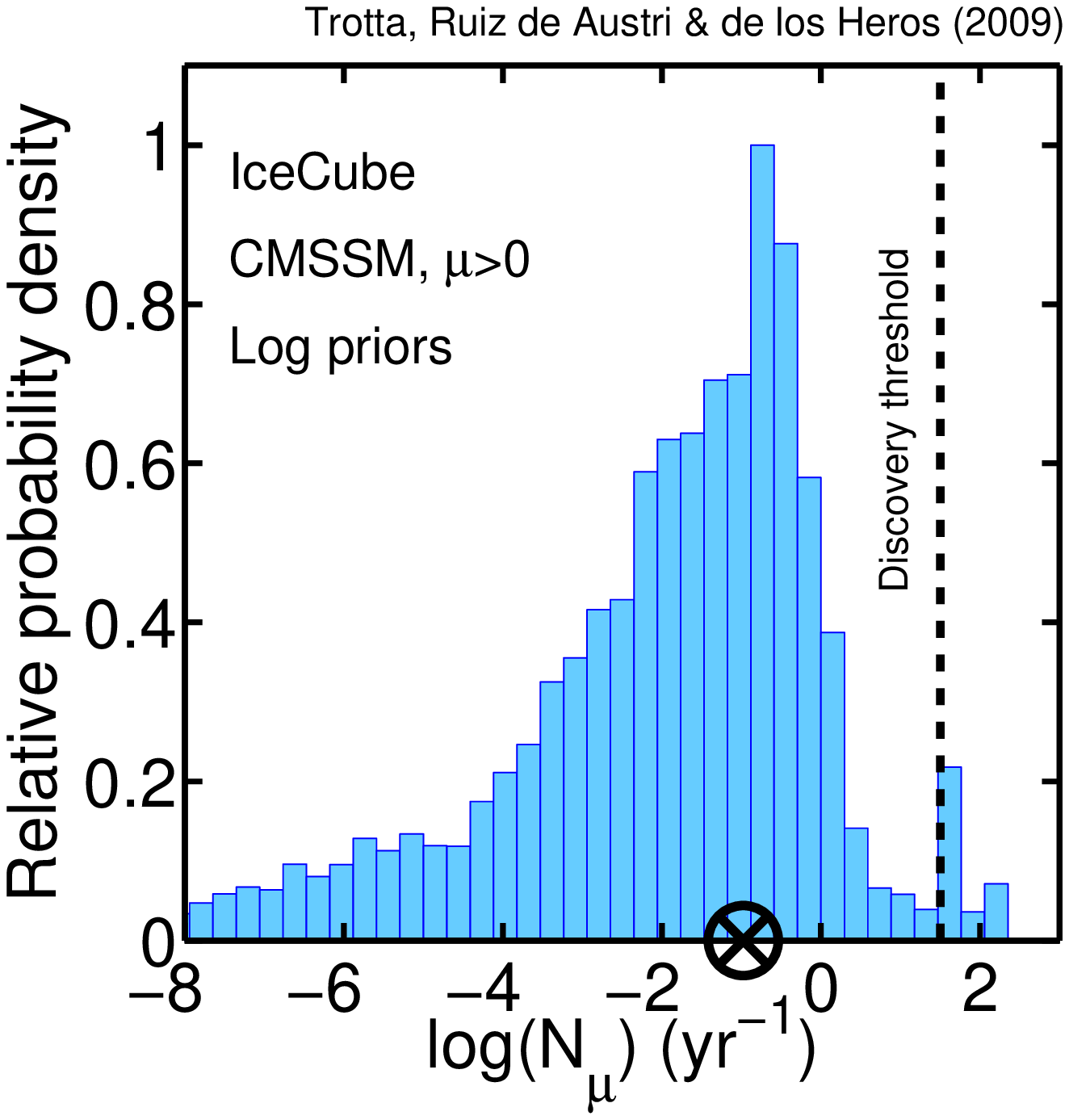}
\caption[test]{Left panel: 2D probability distribution for the number of muon events from the the Sun for IceCube and neutralino mass in the CMSSM, after all available constraints from current data have 
been applied, with 68\% and 95\% contours.  Right panel: 1D probability distribution for the 
number of muon events, with all other parameters marginalised over. The encircled black cross is the best fit point. Both plots assume an NFW profile with $\rholoc = 0.3$ GeV/cm$^3$ 
and $\bar{v} = 270$ km/s for the WIMPs' velocity distribution.}
\label{fig:muevents}
\end{center}
\end{figure}

\begin{figure}[t!]
\begin{center}
\includegraphics[width=\ww]{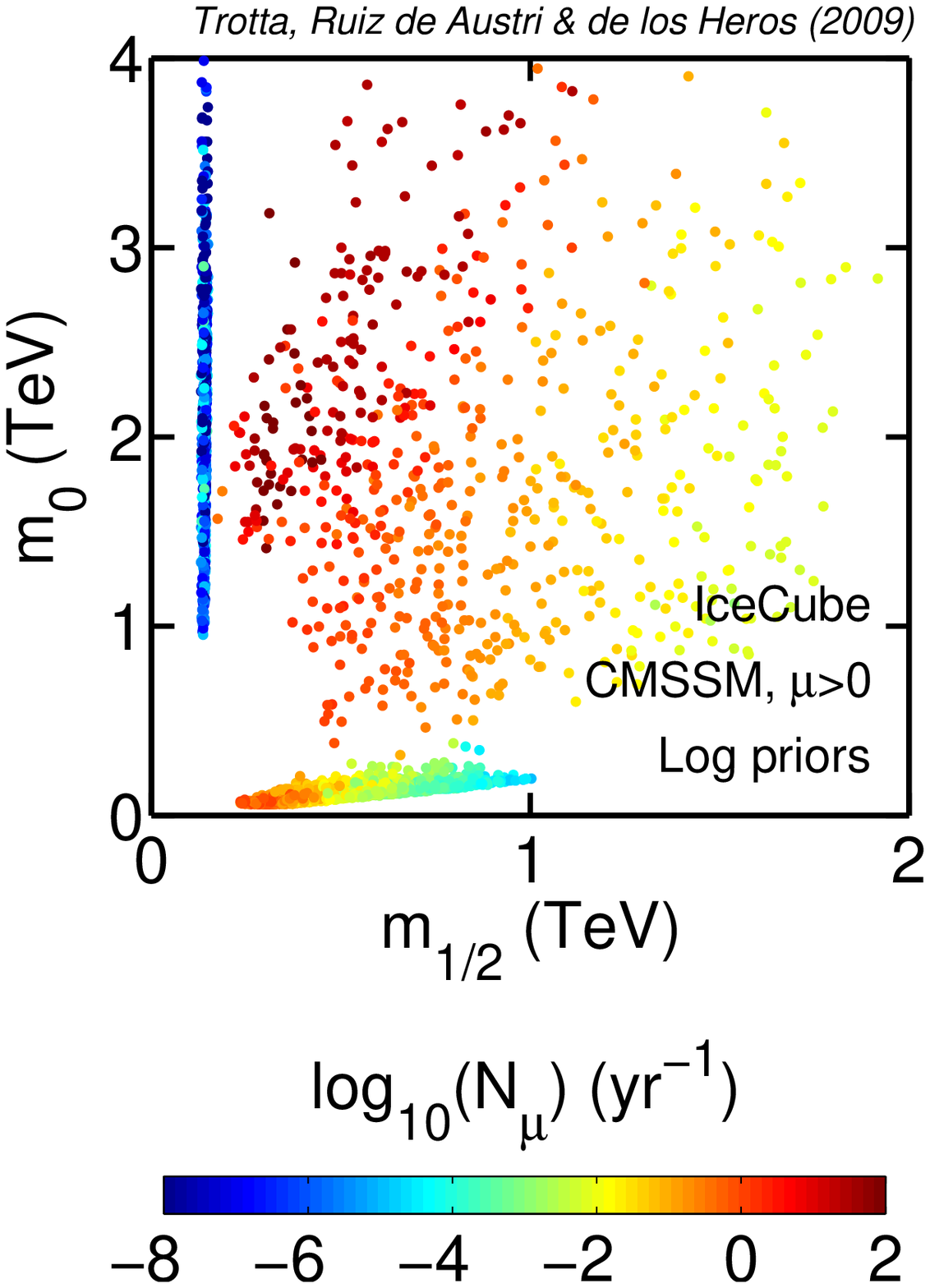}
\includegraphics[width=\ww]{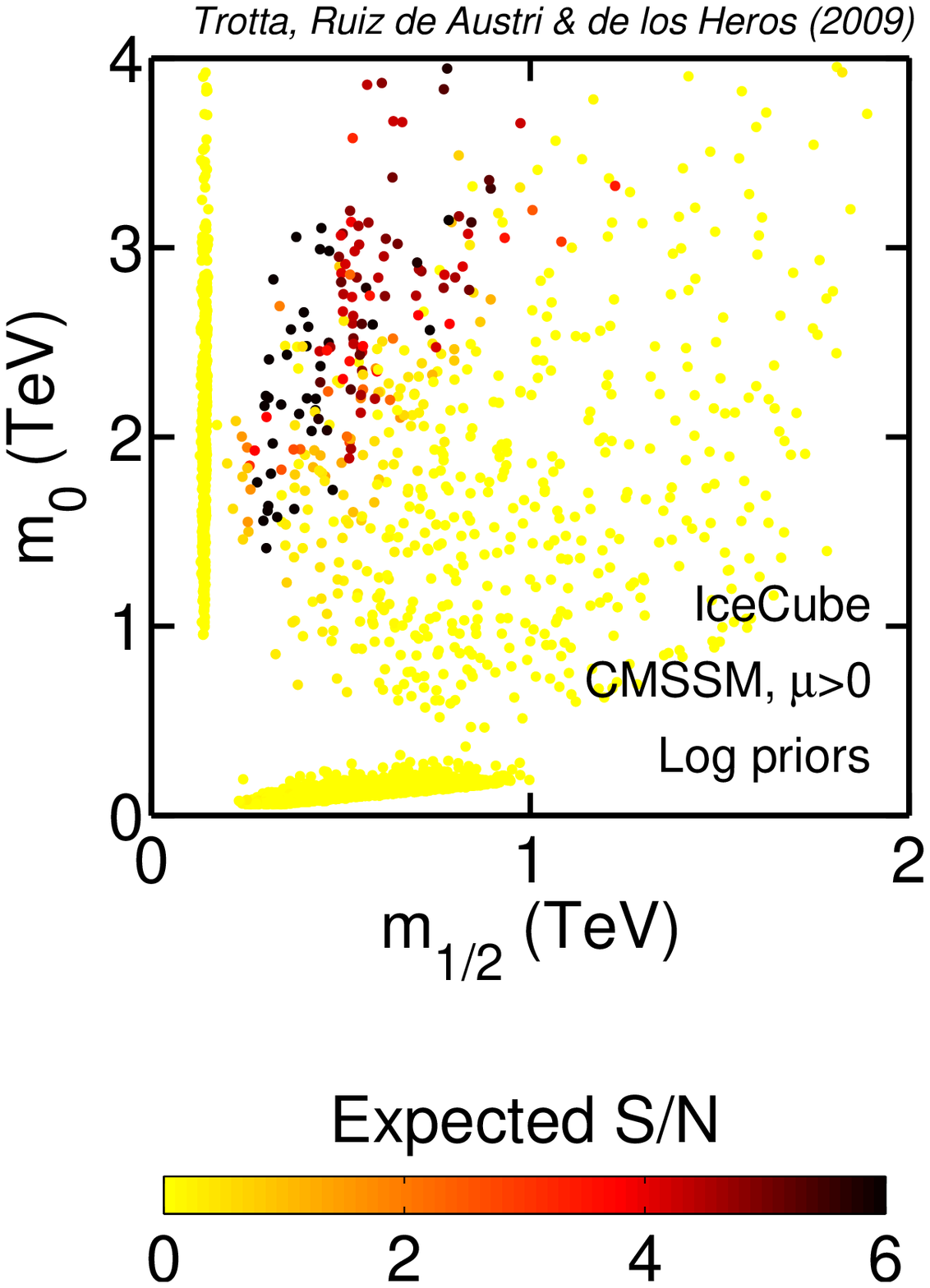}
\caption[test]{Expected number of neutrino--induced muon events from the Sun for IceCube in the $(\mhalf, \mzero)$ plane (left panel, colour coding) and corresponding number 
of $\sigma$ detection significance over the background (right panel). IceCube will be able to probe the parameter space region coloured in dark 
(above the $5\sigma$ detection threshold). The density of samples in each plot reflects probability density.}\label{fig:events_and_sigma_3D}
\end{center}
\end{figure}

We begin by plotting in Fig.~\ref{fig:muevents} the probability distribution of the number of neutrino--induced muon events expected for IceCube and as a function of the neutralino mass, after all available current constraints 
given in Tables~\ref{tab:obs} and \ref{tab:nuisance} have been applied. The posterior probability distribution shows at 95\% an 
island with several hundreds of events, above the discovery threshold. However, most of the probability density favours region of parameter space leading to very few 
events per year, typically $\ll 1$. This is highlighted by the right panel of Fig.~\ref{fig:muevents}, where we plot the 1D marginalised distribution for $N_\mu$. 
The probability density peaks at $N_\mu \sim 0.1 \iyr$, with a long tail extending until $\sim 10^{-10}$ events per year. From this, we find 
that $N_\mu > 2.3\cdot 10^{-7}$ at 95\% (1--tail limit). Further insight is gained by considering the left panel of Fig.~\ref{fig:events_and_sigma_3D}, where we plot the expected number of events in the $(\mhalf, \mzero)$ plane. We can see that the $h$--pole region (thin, vertical sliver at constant $\mhalf$) corresponds to the vertical band in Fig.~\ref{fig:muevents} 
predicting an extremely small number of events (and extending all the way down to $\Nmu \sim 10^{-10}~\iyr$, not shown in the figure).  This is due to the fact that in the $h$--pole region the neutralino mass is low ($\sim 50$ GeV), hence the neutrino spectrum is quite soft and this yields muons with energies mostly below the IceCube threshold. This exemplifies the importance of lowering the energy threshold of neutrino telescopes as much as possible in order to increase the sensitivity to all regions of the CMSSM. The stau-coannihilation region  (the island at small $\mzero$ in Fig.~\ref{fig:events_and_sigma_3D}, corresponding roughly to the 68\%, diagonal region in the left panel of Fig.~\ref{fig:muevents}) also produces a very small number of events.
In this region, the neutralinos are bino--like and they annihilate to 
fermion-anti-fermion pairs, mostly via sfermion exchange. Because the neutralino is a Majorana fermion its
  annihilation cross section to fermion pairs is s-wave suppressed
  and $\langle \sigma_a v \rangle \propto m_f^2/m_\chi^2$. As a consequence, the contribution to the total cross section from $\bbar$ and $\tau^+ \tau^-$ channels is small. Therefore the annihilation rate and capture rate are far from being in equilibrium (see section~\ref{sec:equilibrium} for a more thorough discussion), and the ensuing neutrino flux is small. A special role is played by the top quark since, if this channel is kinematically allowed, one expects a larger cross section and hence a larger neutrino flux due to its heavier mass. This requires both a larger value of the neutralino mass and that the neutralino be a mixture state of bino and higgsino. Both conditions are satisfied in the focus point  region at large $(\mhalf, \mzero)$, which typically corresponds to a larger number of events per year, of the order of up to several hundreds.

We can convert the probability distribution for $N_\mu$ into the probability distribution for a detection 
at a given S/N level, estimated according to Eq.~\eqref{eq:SN}.  The distribution of the significance level in the $(\mhalf, \mzero)$ plane is shown in the right panel of Fig.~\ref{fig:events_and_sigma_3D}, where the black points have fluxes above the $5\sigma$ detection threshold for IceCube. This demonstrate that IceCube will be able to probe at high significance the focus point region of the CMSSM.  We find that IceCube has a 1.7\% (3.4\%) probability of obtaining a $5\sigma$ ($3\sigma$) detection of WIMPs from the Sun. However, a large fraction of the CMSSM parameter space favoured by current data will remain unaccessible to IceCube. It has to be kept in mind that the relative importance of the focus point is difficult to assess with present--day data. In fact, a different choice of priors for the CMSSM parameters leads to a significant increase in the probability of the focus point, and hence to more optimistic detection prospects for IceCube. We comment on the impact of the change of CMSSM priors in section~\ref{sec:prior_change}, where we also compare our results with previous studies. 

A modification in our astrophysical assumptions and/or in the choice of data employed to constrain the CMSSM might lead to modify somewhat the above results. It has been suggested~\cite{Read:2009iv} that the Milky Way might also host a ``dark disk'' (additionally to the dark matter halo) with a population of co-rotating WIMPs whose capture rate for the Sun could be up to 1 order of magnitude larger than for halo WIMPs due to their different velocity distribution~\cite{Bruch:2009rp}. We can qualitatively estimate the effect of such a dark disk by considering an increase in the muon flux by a factor of 10. This would increase the probability of a $5\sigma$ detection to 4.4\%. From the particle physics side, it is worth emphasizing that the pull towards the low--mass coannihilation region comes, to a large extent, from one single observable, namely the anomalous magnetic moment of the muon, for which a discrepancy with the SM value persists at the $\sim 3 \sigma$ level. However, if one removes this somewhat controversial observable from the analysis, the relative importance of the focus point region (and therefore the detection prospects for IceCube) might increase significantly~\cite{tfhrr:2008}.  

\subsection{Contribution from different annihilation channels}

We now turn to investigating the relative contribution of the different annihilations channels to the signal for IceCube. In model--independent analyses of present--day neutrino telescope 
data, it is often assumed that the neutrino spectrum is dominated by one of the 29 available final states. Typically, the analysis is carried out for a soft neutrino 
spectrum, produced in quark jets from e.g. $\bbar$, and a harder spectrum, ensuing from the decay of $\tau$ leptons or gauge bosons  (e.g.,~\cite{Ackermann:2005fr,IceCube:09a}). 
It is then argued that such choices will roughly cover the phenomenology of ``typical'' points in parameter space. There are two potential issues with this procedure. First, once an underlying model is specified (here the CMSSM), 
the relative importance of soft and hard channels will generally vary across the parameter space, and in a correlated way. Therefore, limits obtained assuming that only one of the channels dominates in the whole parameter space cannot be automatically applied to a model--specific scenario. Secondly, even in a purely phenomenological approach, incorrectly assuming that one of the channels dominates while the signal really comes from a different final state or from a mixture of states leads to potentially large systematic errors in the WIMP 
parameters extracted from indirect detection methods. This has been demonstrated in the case of $\gamma$ 
ray observatories~\cite{Jeltema:2008hf}, and, as we show below, the same {\em caveat} applies in the context of neutrino telescope data.   

\subsubsection*{Benchmark points}
\begin{figure}[t!]
\begin{center}
\includegraphics[width=\ww]{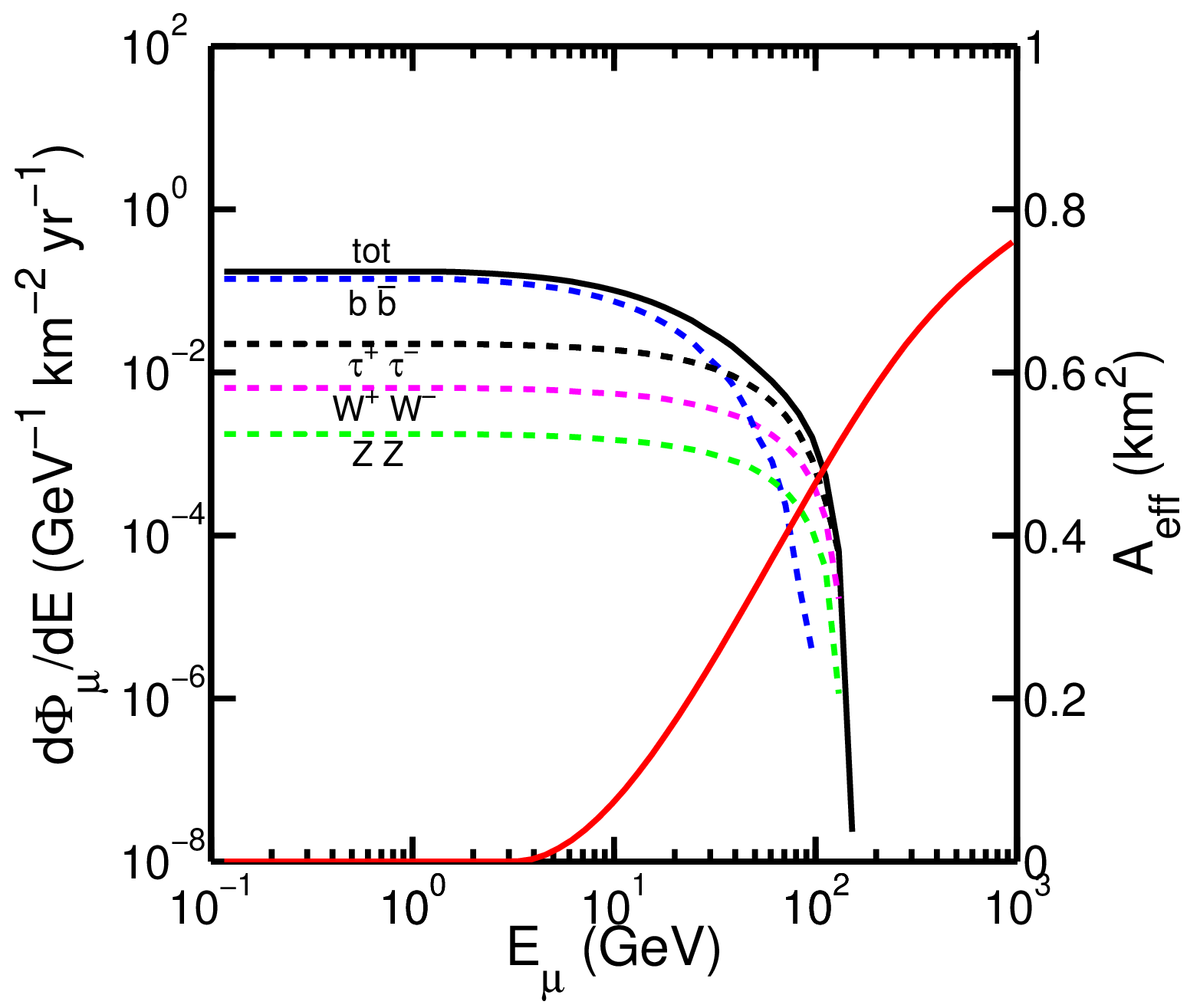}
\includegraphics[width=\ww]{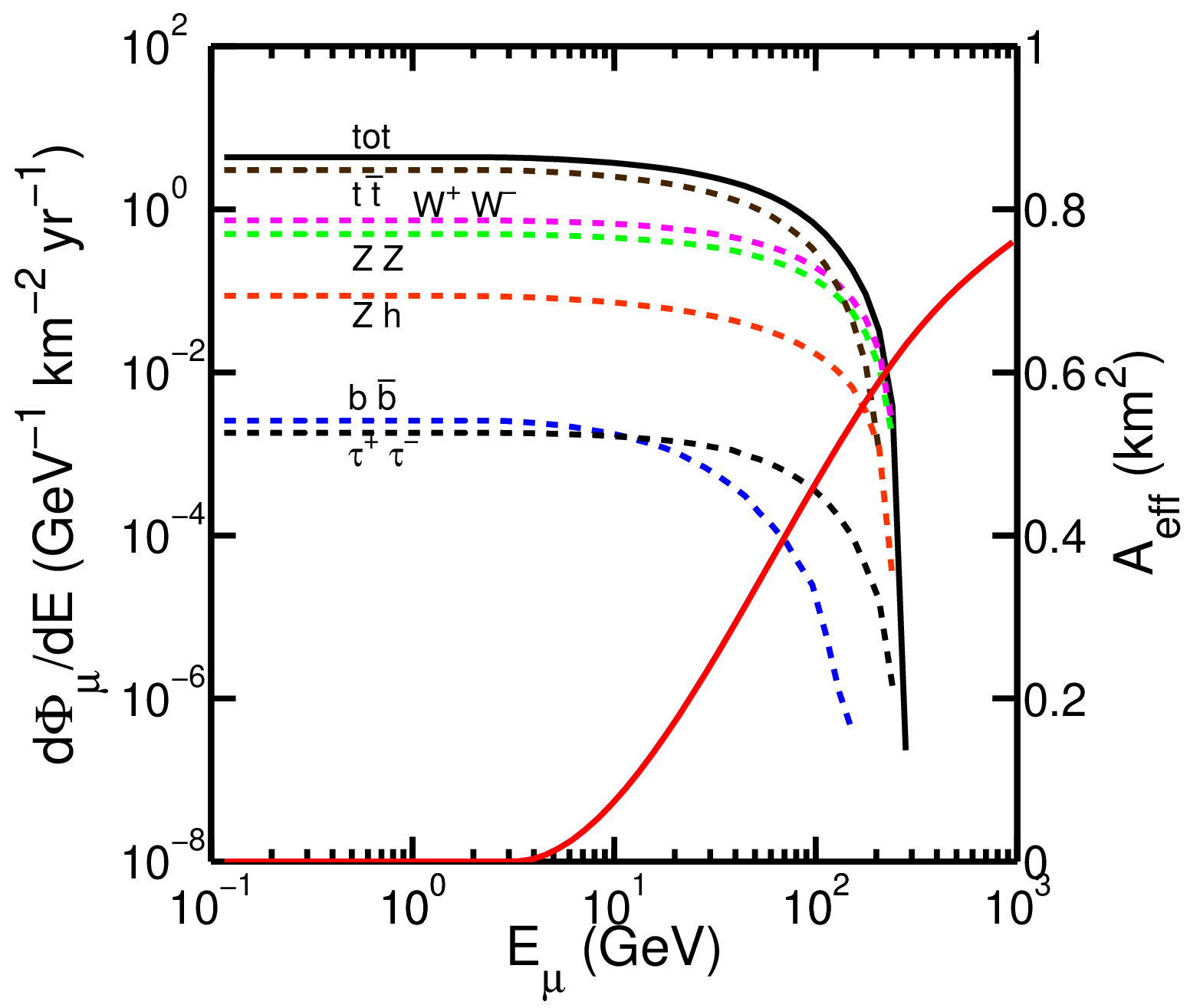}
\caption[test]{Differential muon flux for the dominant annihilation channels for benchmark point A (left panel) and benchmark point B (right panel). The solid, black is the total spectrum, while the solid red line depicts the effective area used in the convolution. }

\label{fig:diff_flux}
\end{center}
\end{figure}

Figure~\ref{fig:diff_flux} shows the differential muon flux for two benchmark points, depicting the contribution from the most important channels. The left panel (benchmark point A) is for the overall best--fit point from our scan, which lies in the coannihilation region, while the right panel is for a point in the focus point region, within the reach of IceCube (benchmark point B). The parameters values for the two benchmark points are given in Table~\ref{tab:benchmark}, along with their branching ratios and the fractional contribution to the number of muon events for each channel. The branching ratio of benchmark point A is dominated by the $\bbar$ final state (94.7\%), but this channel accounts for only about 8\% of the number of events. This is because the soft $\bbar$ spectrum is mostly in the energy region where the effective area of IceCube is small, see the left panel of Fig.~\ref{fig:diff_flux}. As a consequence, the harder $\tau^+\tau^-$ spectrum, whose branching ratio is only 1.5\%, is responsible for more than half of the expected number of events. Overall, this benchmark point will remain outside the reach of IceCube (although detection prospects would be more favourable for a DeepCore configuration), as the total number of events is $N_\mu = 0.11$ per year. Benchmark point B is rather different, as the two most important channels in terms of branching ratio are $\chi\chi \rightarrow \bar{t}t$ (71\%) and final states involving gauge bosons (16.6\% for $W^+W^-$ and 9.4\% for $ZZ$). After convolution with the effective area (see right panel of Fig.~\ref{fig:diff_flux}), the relative importance of those channels changes very noticeably, with the final state involving $\tbar$ contributing about the same fraction of events as the final states involving gauge bosons together.

\begin{table}[t]
\centering
\begin{tabular}{| l | l l |}
\hline
Quantity &  Benchmark A &  Benchmark B   \\\hline
\multicolumn{3}{|c|}{CMSSM and SM parameters} \\\hline
$\mzero$ (GeV)  &  125.6 & 2809.6\\ 
$\mhalf$ (GeV)  &   372.9 & 666.6\\
$A_0$  (GeV) & 728.6 & 719.2\\
$\tanb $   & 18.41 & 12.13\\
$\mtpole$ (GeV)  & 173.6 & 171.0 \\
$m_b (m_b)^{\overline{MS}}$ (GeV) & 4.18 & 4.34  \\
$\alphas$ &    0.1179 & 0.1170  \\ 
$1/\alphaemmz$    &     127.96 & 127.90 \\\hline
\\ \hline
\multicolumn{3}{|c|}{Cosmological relic abundance} \\\hline
$\Omega_\chi h^2$ & 0.1145 & 0.1022 \\ \hline
\multicolumn{3}{|c|}{Neutralino mass} \\\hline
$m_\chi$ (GeV) & 148.2 & 267.2 \\ \hline
\multicolumn{3}{|c|}{BR's for dominant channels} \\\hline
$\text{BR}(\chi \chi \rightarrow b\bar{b}$) & 0.947 & $2.1\times10^{-3}$ \\     
$\text{BR}(\chi \chi \rightarrow t\bar{t}$) & 0 & 0.717 \\
$\text{BR}(\chi \chi \rightarrow \tau^+\tau^-$) & 0.015 & $1.6\times10^{-4}$  \\
$\text{BR}(\chi \chi \rightarrow W^+W^-$) & 0.011 & 0.162  \\
$\text{BR}(\chi \chi \rightarrow ZZ$) & $2.5\times10^{-3}$ & 0.094  \\ \hline
\multicolumn{3}{|c|}{Muon events for IceCube} \\\hline
$\Nmu \iyr$ (total) &  0.11 & 34.8\\
$N_{ b\bar{b}}/\Nmu$ & 0.08 & $5\times10^{-5}$ \\
$N_{ t\bar{t}}/\Nmu$ & 0 & $0.46$ \\
$N_{ \tau^+\tau^-}/\Nmu$  & 0.56 & $5\times10^{-4}$ \\
$N_{ W^+W^-}/\Nmu$ & 0.24 &  0.30\\
$N_{ ZZ}/\Nmu$ & 0.06 & 0.21\\\hline
\end{tabular}
\caption{Values of the CMSSM and SM parameters and corresponding branching ratios as well as number of muon events in each channel for the two benchmark points discussed in the text. Benchmark point A is the overall best--fit for the CMSSM, and it lies in the coannihilation region. Benchmark point B is in the focus point region (within $\sim 2\sigma$ of the best--fit) and would be detectable with $\sim 5\sigma$ significance by IceCube. We also give the value of the cosmological relic abundance and of the WIMP mass for completeness. \label{tab:benchmark}}
\end{table}

\subsubsection*{Systematic error from assuming a single dominating channel}
   
Let us now consider more in detail benchmark point B, for this represents a fiducial 
model that would be detectable with $\sim 5\sigma$ significance by IceCube. If one assumed (as it is done in current data analysis from 
neutrino telescopes) that the spectrum is dominated by one single channel (i.e., assuming a branching ratio of 1 for that channel), one would introduce a serious systematic error in the 
analysis. As an illustration, if we assumed that the $\bbar$ was the dominant channel, then we would predict $N_\mu \sim 0.8$ events 
per year, rather than the correct value $N_\mu = 34.8 \iyr$, thus underpredicting the true rate by a factor $\sim 40 $. If instead 
one considered the $W^+W^-$ channel only, then one would obtain $N_\mu \sim 64 \iyr$, thereby overpredicting the number of events 
by a factor $\sim 2$. In current practice, the observed number of events is used to derive an upper limit to the number of events 
due to a possible WIMP signal. This is then translated into a corresponding limit for the muon flux, assuming one dominating channel. 
This step can potentially introduce a large systematic error in the flux limit, as the above example illustrates, arising from making 
the extreme assumption of annihilation into only one dominant channel (when the actual signal is coming from several channels). 
Since the flux constrained by the experiment is the product of the branching ratio times the annihilation cross section, this 
systematic error would propagate to the estimation of the annihilation cross section and from there to the scattering cross section 
(derived under the assumption of equilibrium). This is acceptable when data only give an upper limit, as long as one keeps in mind 
that the limits derived under a typical soft (hard) channel are going to be an underestimate (overestimate) of the true limits. However, 
this problem becomes a source of concern once a signal is detected, for in that case it is clearly desirable to avoid introducing 
large systematic errors in the estimates of the annihilation cross section and of the neutralino mass (the latter effect coming from 
employing the wrong spectral shape when one assumes a single dominating channel). A methodology that accounts for this and avoids 
introducing systematic biases will be presented elsewhere.

\begin{figure}[t!]
\begin{center}
\includegraphics[width=\qq]{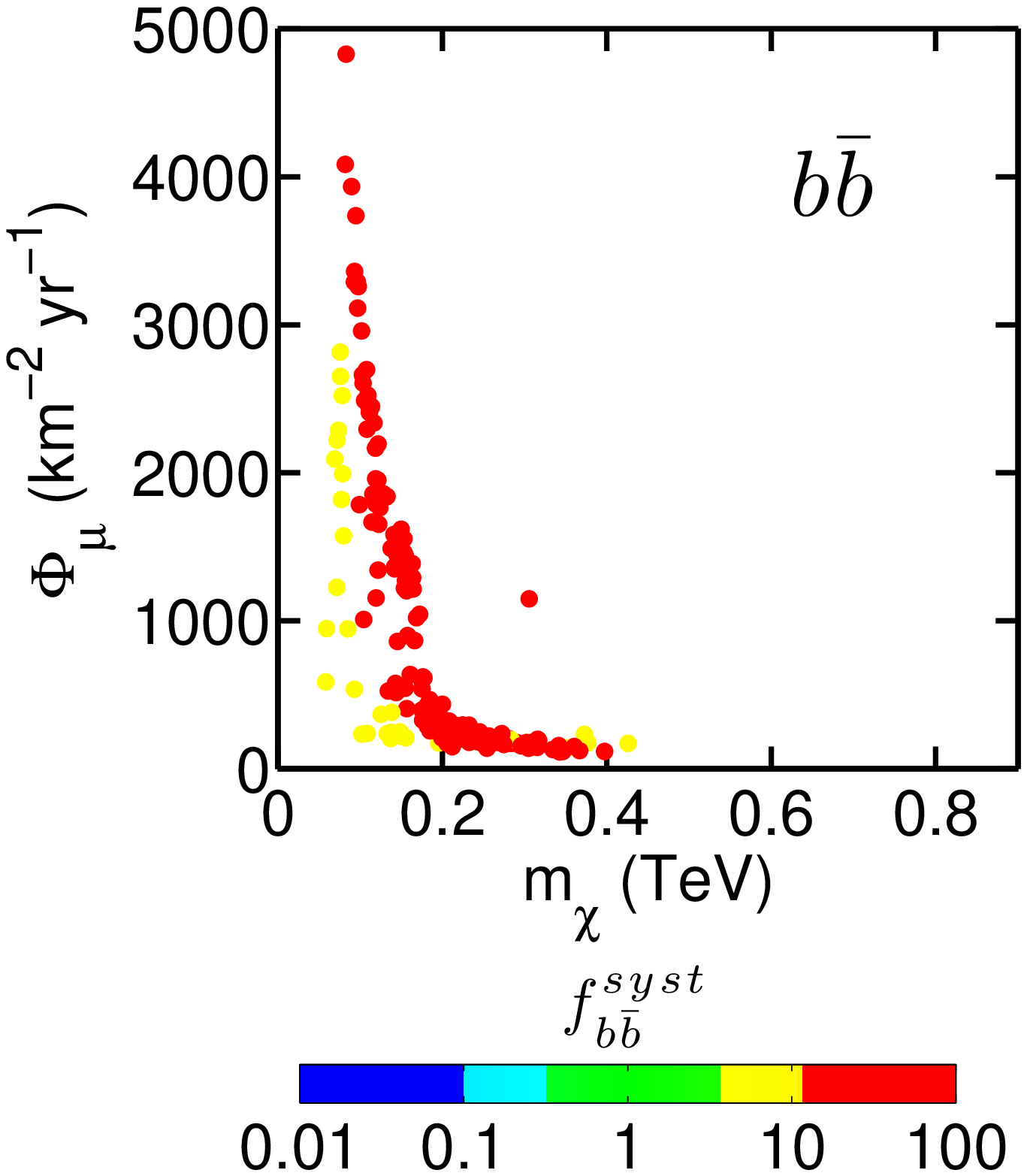}\hfill
\includegraphics[width=\qq]{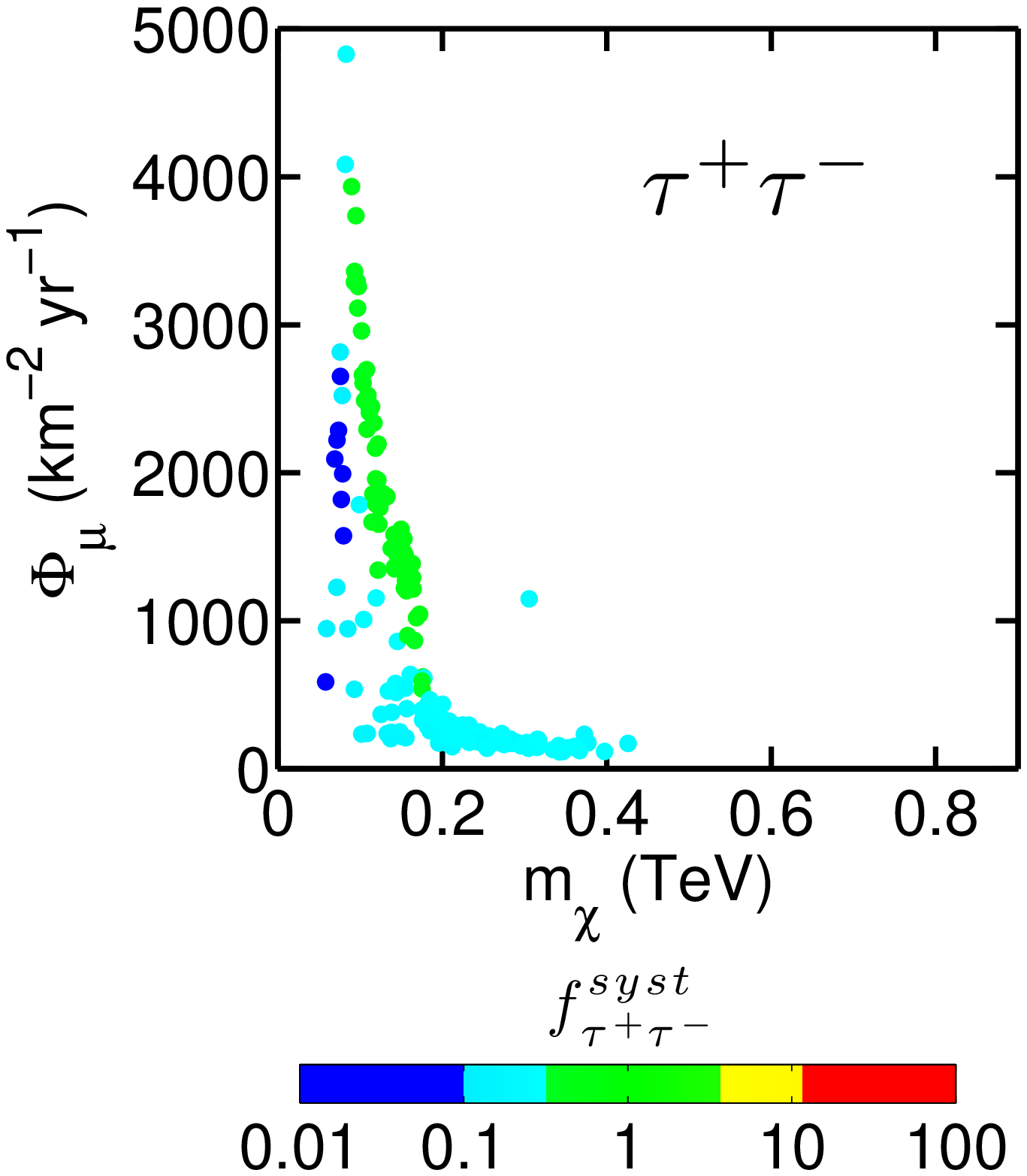}\hfill
\includegraphics[width=\qq]{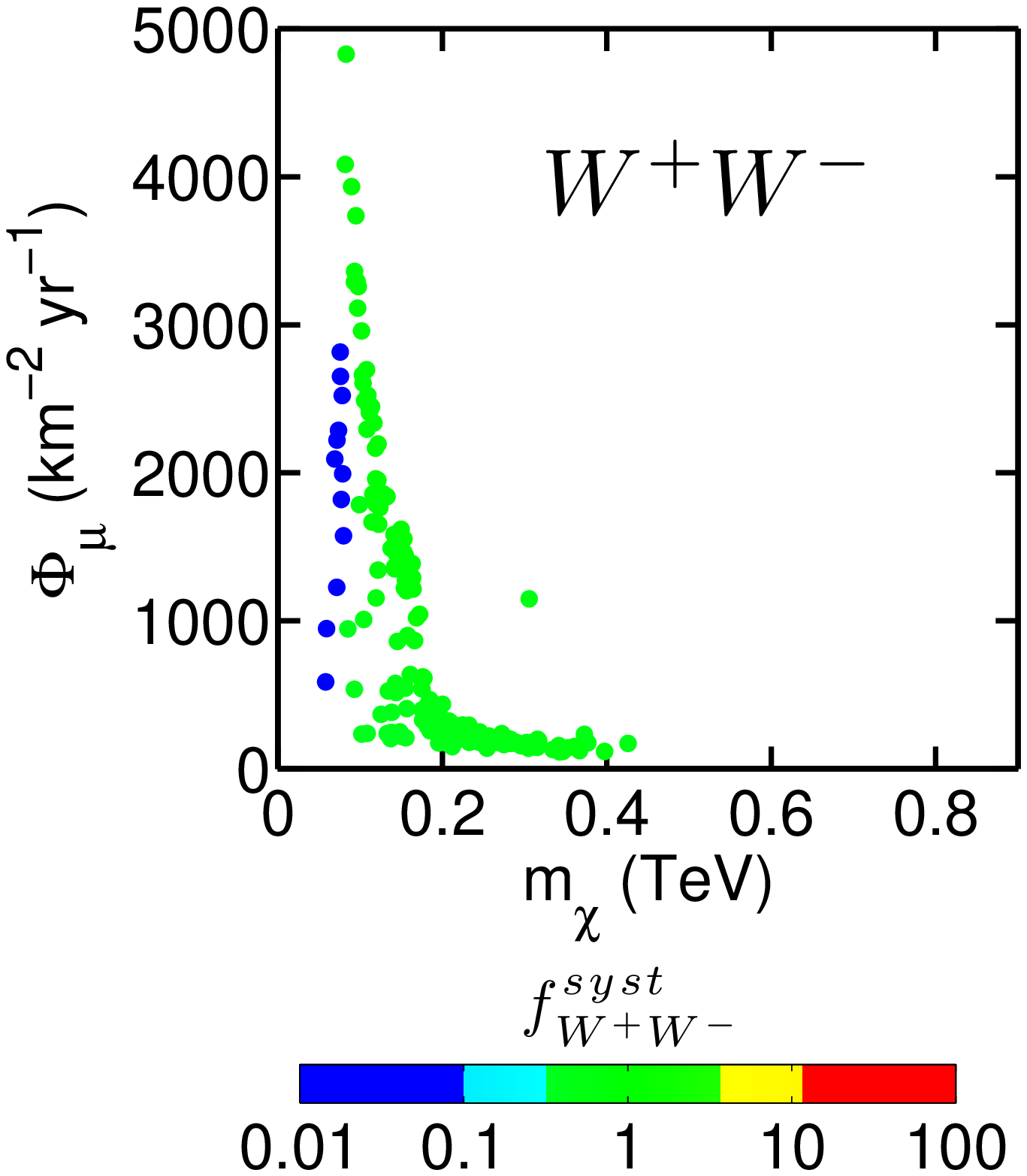}\hfill
\includegraphics[width=\qq]{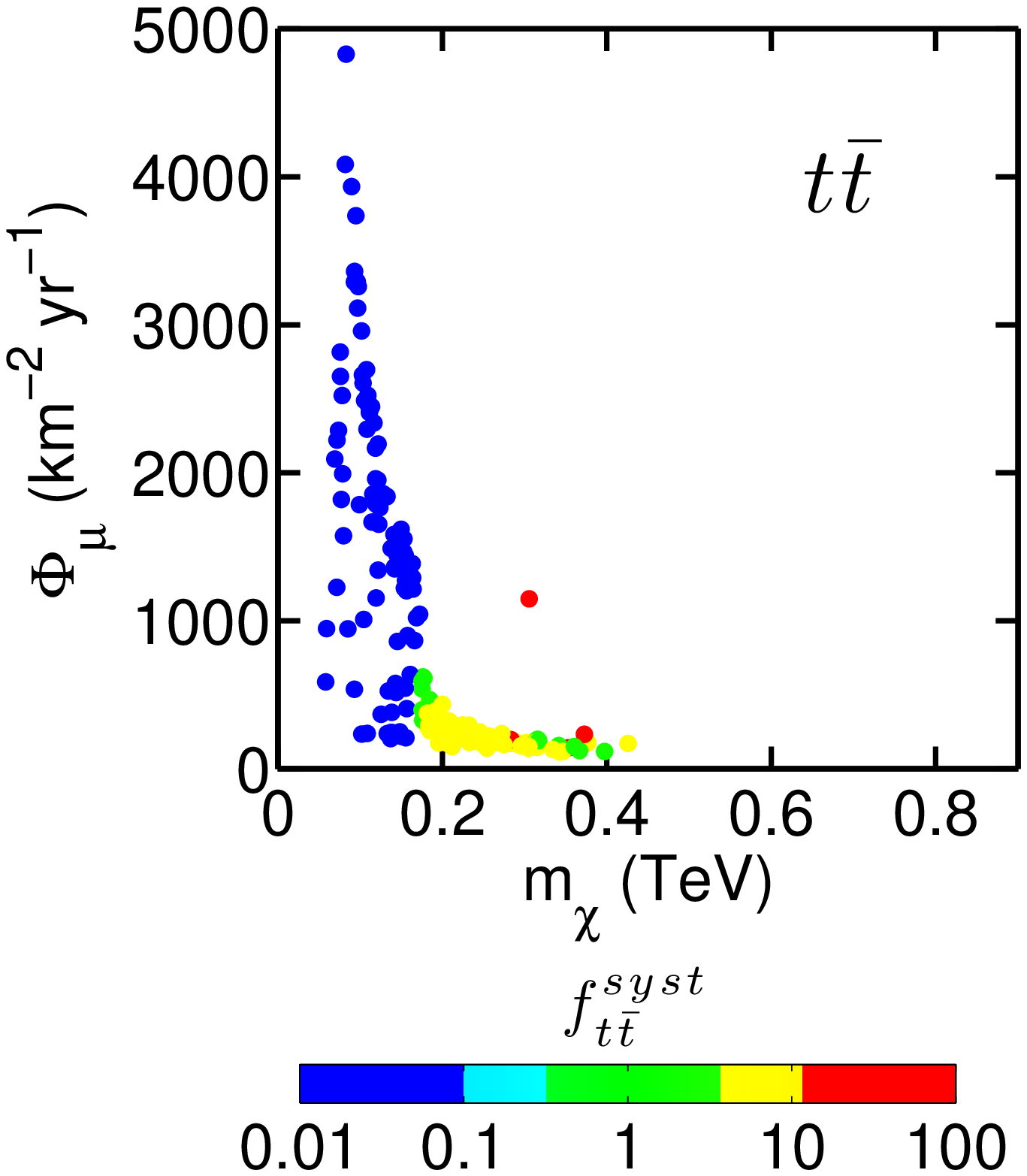}
\caption[test]{Systematic error factor induced in the reconstructed annihilation rate from (incorrectly) assuming that the signal is 
dominated exclusively by one of the four channels shown (from left to right: $\bbar$ , $\tau^+\tau^-$, $W^+W^-$ and  $\tbar$), as a 
function of the muon flux above 1 GeV and the neutralino mass (notice that the muon flux has been converted to the conventional 1 GeV threshold, and that the scale of the vertical axis is 
linear). Green points are the least biased, having a systematic error below a 
factor of 3. We show only samples within the discovery region of IceCube.}
\label{fig:syst_Phimu_mchi}
\end{center}
\end{figure}

In general, we can introduce a factor $\fsyst_i$ which measures the error introduced in the annihilation and scattering cross section 
estimation when assuming that a single channel $i$ dominates the annihilation (assigning it a branching ratio of unity) 
\be
\fsyst_i = \frac{\text{BR}(\chi \chi \rightarrow i)}{N_i/N_\mu},
\ee
where N$_i$ represents the number of events due to annihilation channel $i$ and $\text{BR}(\chi \chi \rightarrow i)$ is the true branching ratio for that final state. 
The systematic error is largest for points in parameter space whose contribution to the fractional number of events is very different 
from their branching ratio. Notice that $\fsyst_i \gg 1$ when the branching ratio is close to unity but $N_i/N_\mu \ll 1$, which is 
the case for the soft channels, while $\fsyst_i \ll 1$ when the signal is dominated by one or more channels with a small branching 
ratio (typically the case for $i = W^+,W^-, ZZ$ or $\tau^+\tau^-$). Fig.~\ref{fig:syst_Phimu_mchi} shows the value of $\fsyst_i$ for 
$i=\bbar, \tau^+\tau^-, W^+W^-$ and  $\tbar$ in the plane of the neutralino mass and the muon flux\footnote{For the $\bbar$ final state 
there are several points where this channel contributes no events at all. Such points have been arbitrarily mapped to a value 
$\fsyst_{\bbar} = 100$.}. The figure only shows samples in the CMSSM parameter space in the IceCube discovery region.  The final state which leads to the least bias is $\chi\chi \rightarrow W^+W^-$ (less 
than 50\% error across most of the accessible region), while assuming a $\bbar$ ($\tbar$) final state leads to a systematic 
underestimation (overestimation) of order $\sim 100$ or larger. 

\subsubsection*{Final-state contributions to the signal}  
\begin{figure}[t!]
\begin{center}
\includegraphics[width=\qq]{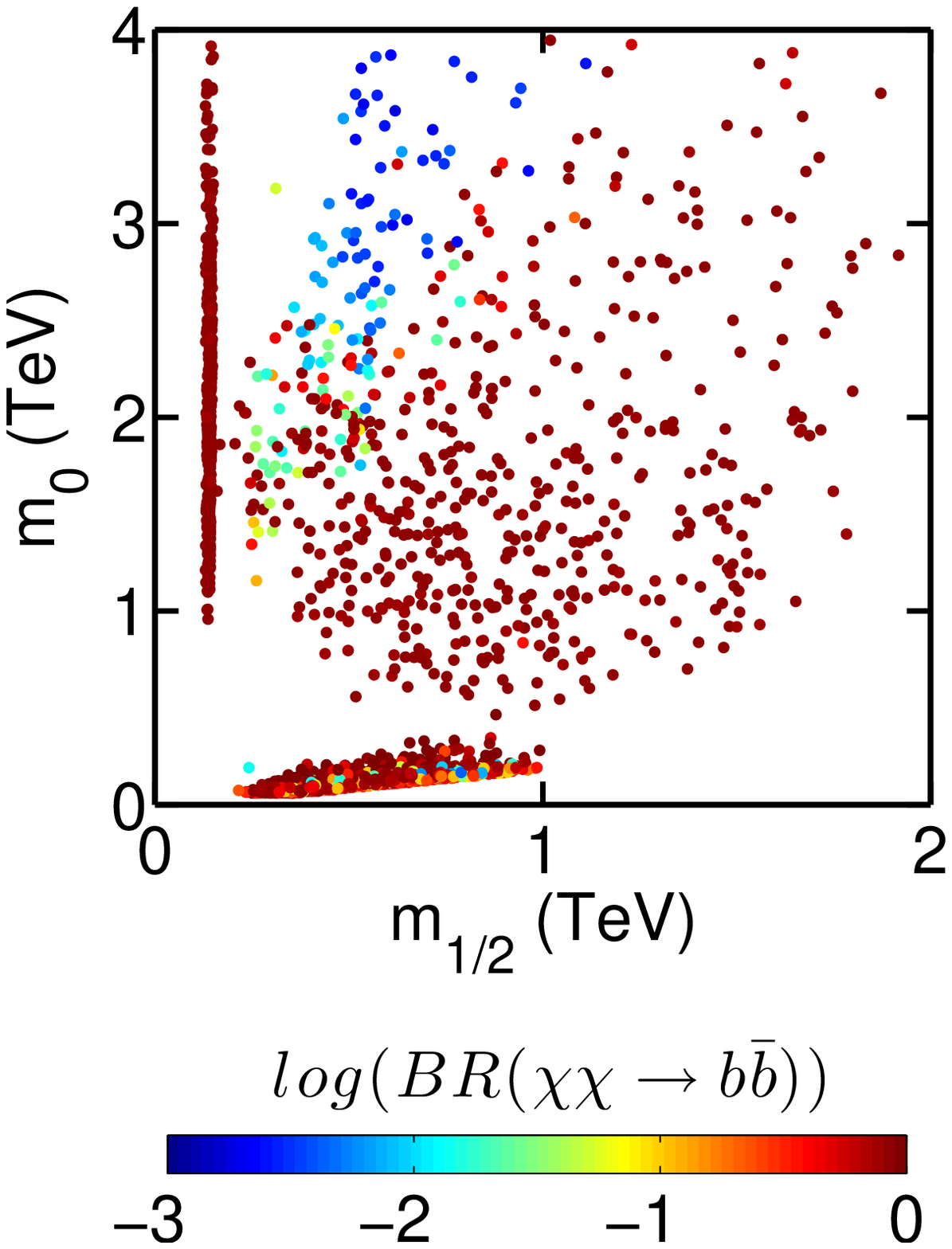}\hfill
\includegraphics[width=\qq]{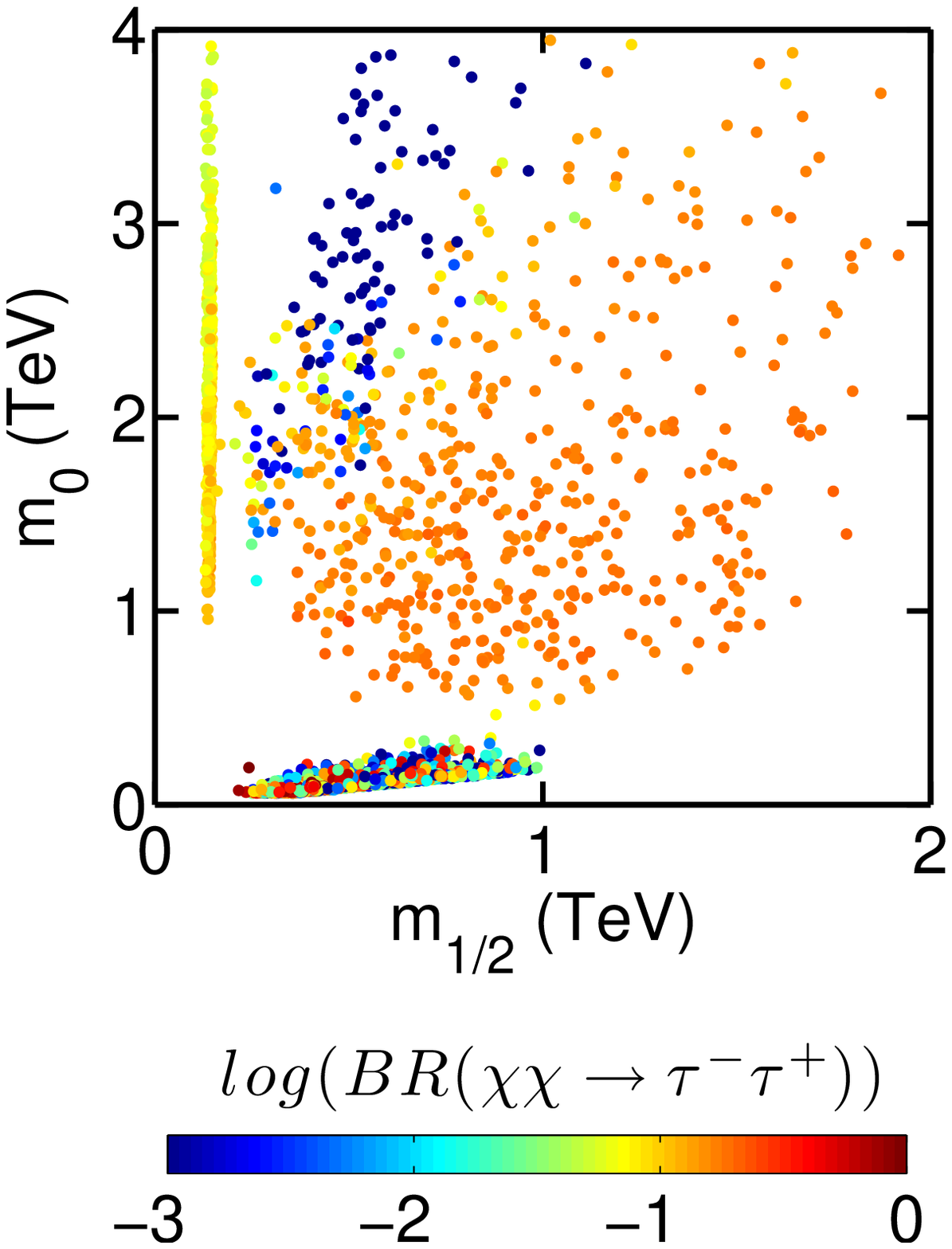}\hfill
\includegraphics[width=\qq]{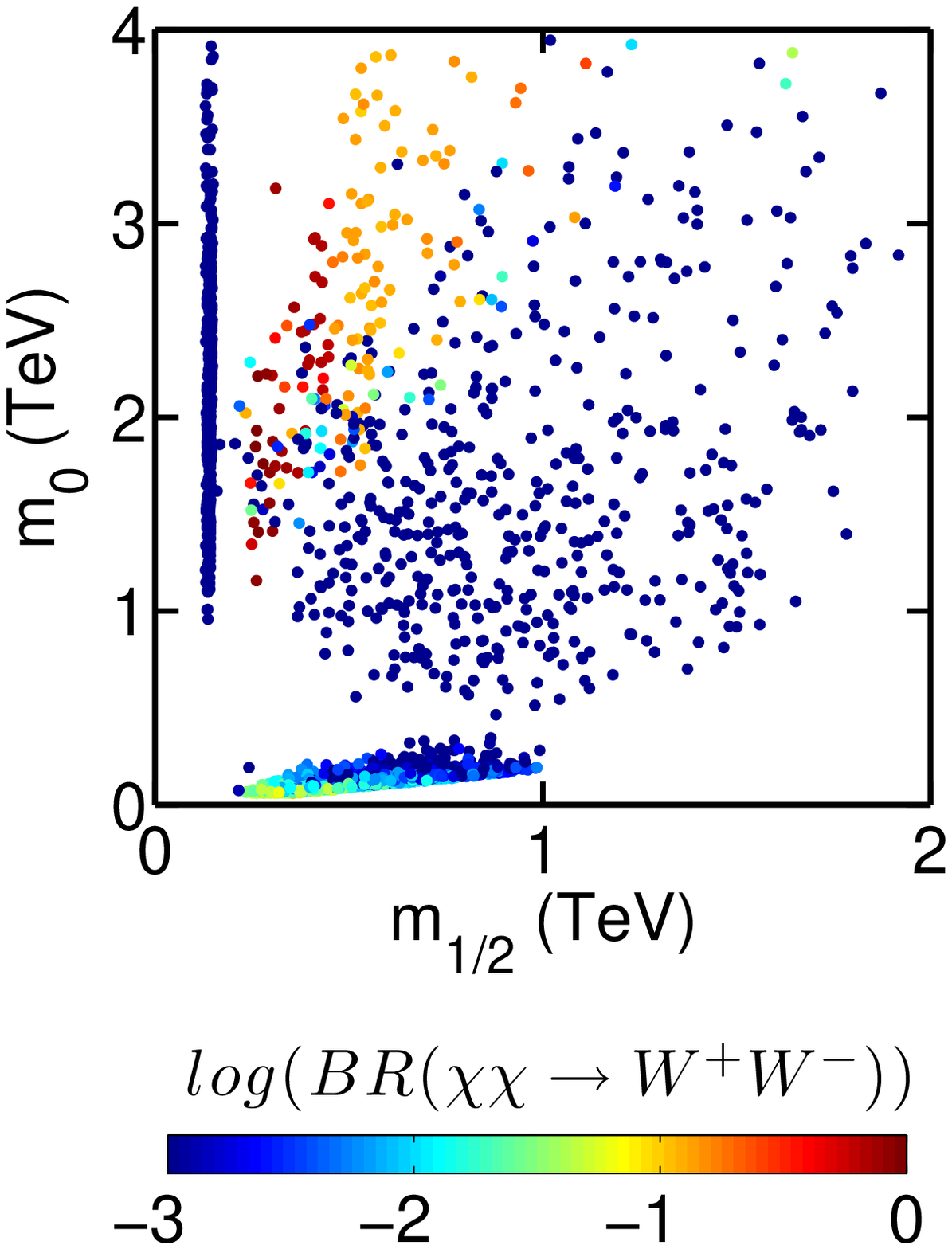}\hfill
\includegraphics[width=\qq]{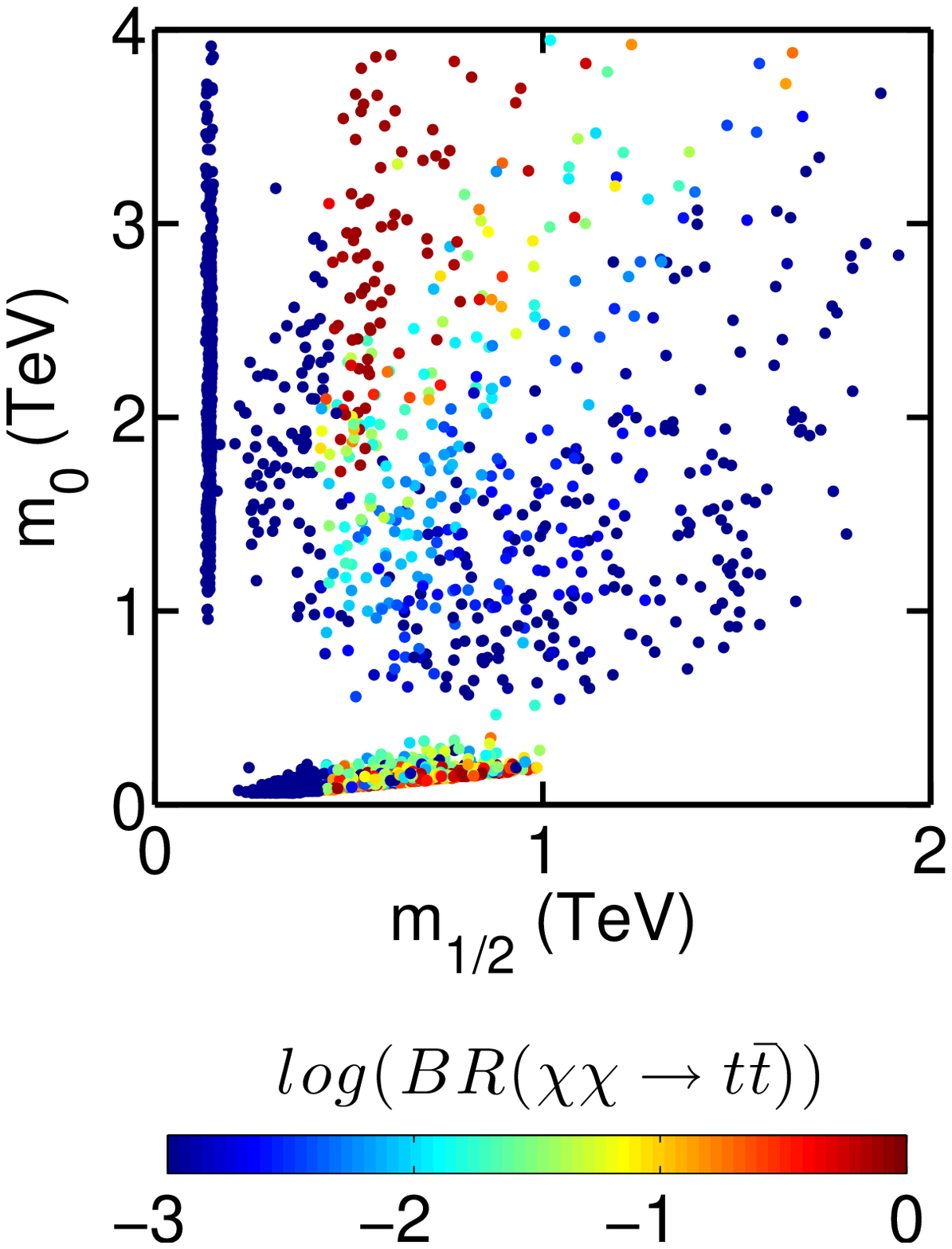}\\
\includegraphics[width=\qq]{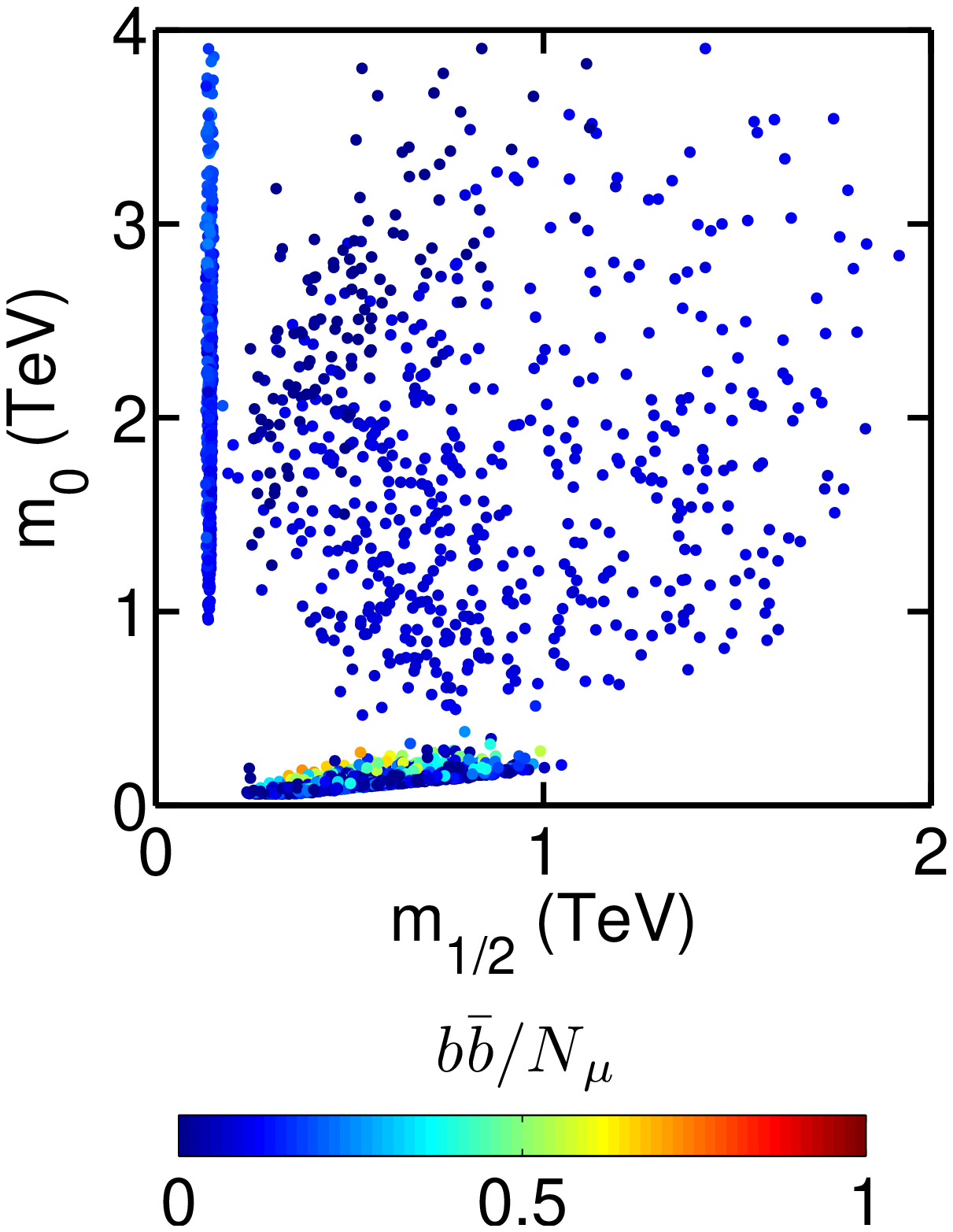}\hfill
\includegraphics[width=\qq]{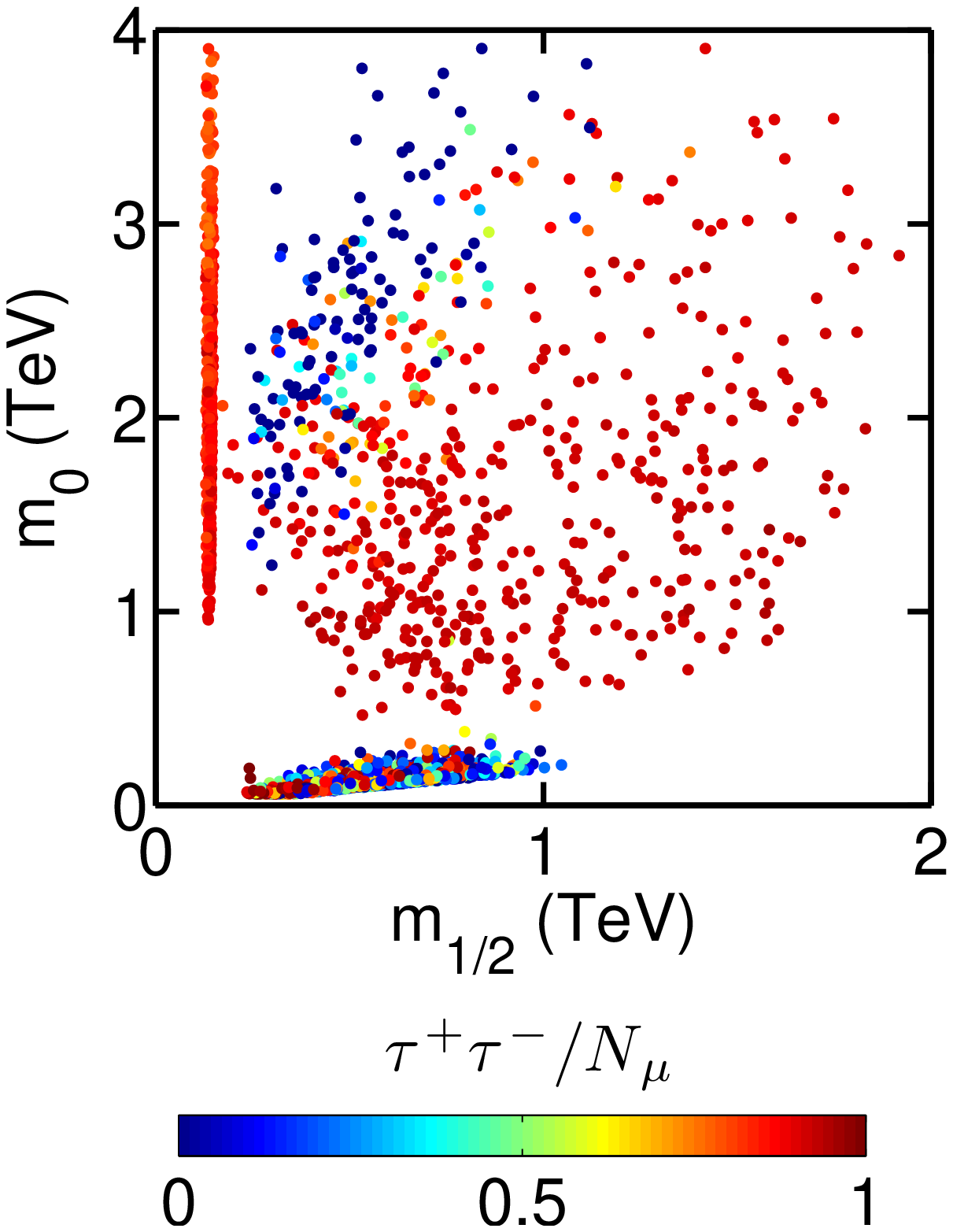}\hfill
\includegraphics[width=\qq]{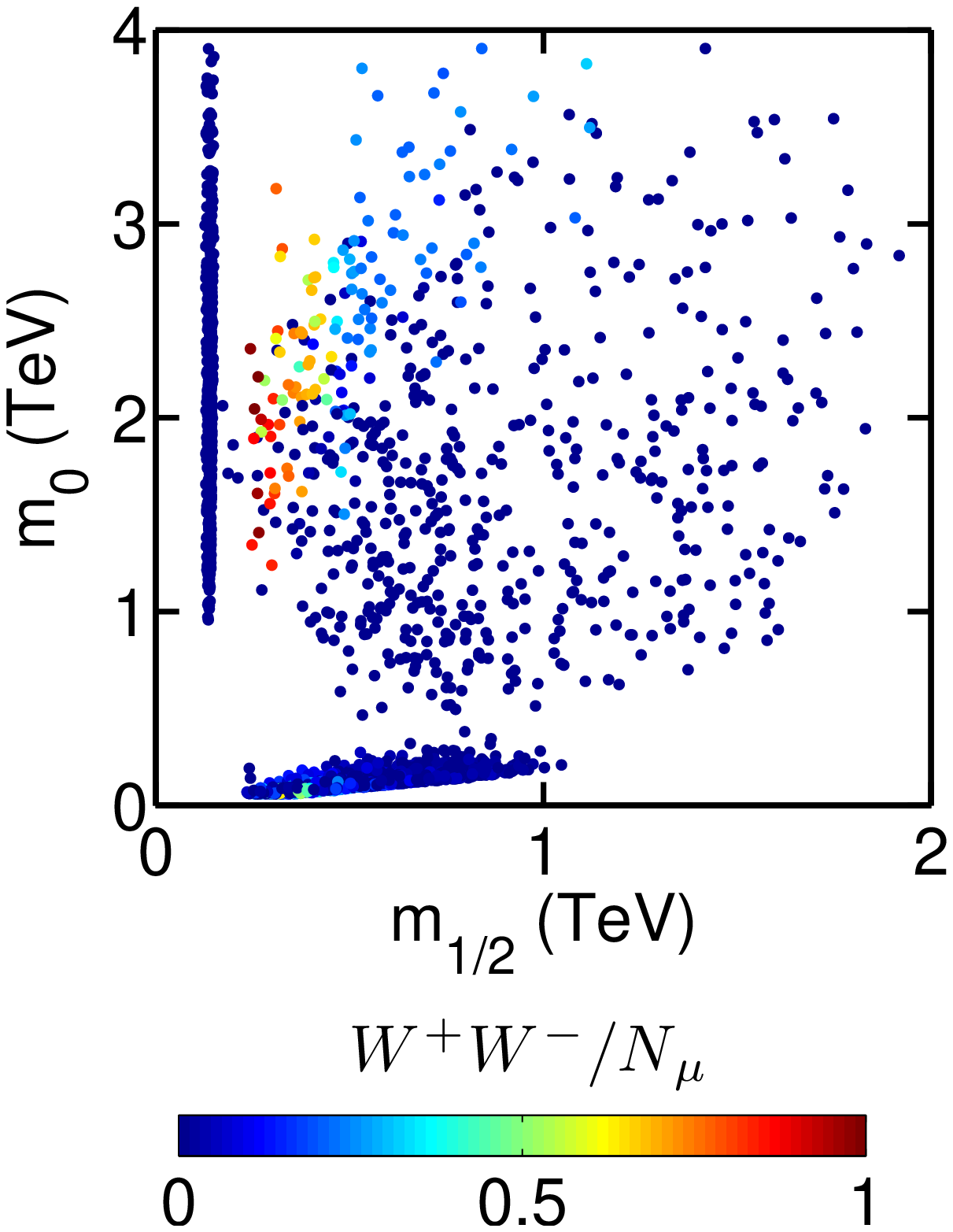}\hfill
\includegraphics[width=\qq]{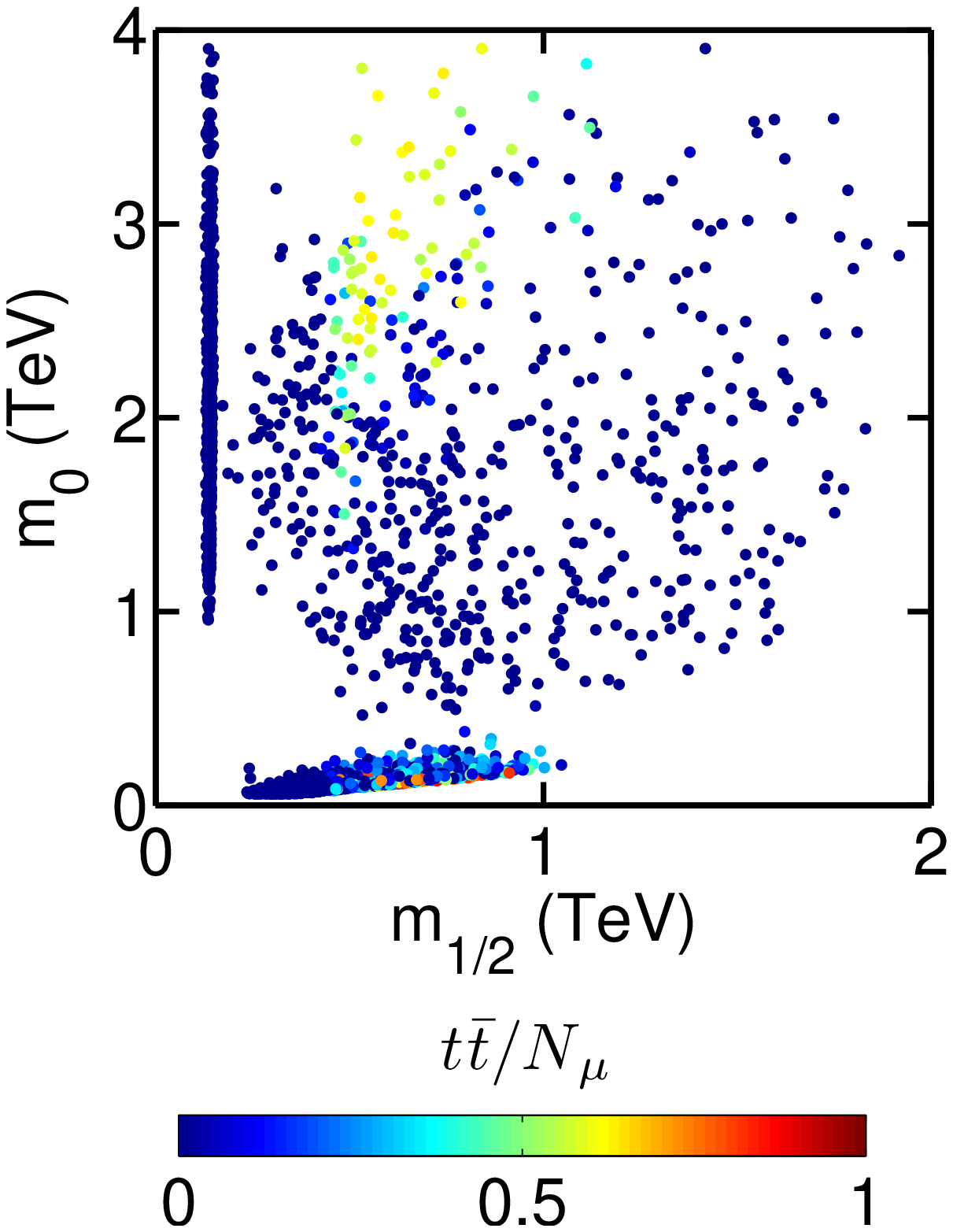}

\caption[test]{Top row: branching ratios in the ($\mhalf, \mzero$) plane for the most important annihilation channels in the CMSSM, 
from left to right:  $\bbar$, $\tau^+\tau^-$, $W^+W^-$ and $\tbar$ (the plots for $ZZ$ are similar to the ones for $W^+W^-$  and are 
not shown). Bottom row: fractional number of muon events expected in IceCube from each channel.}
\label{fig:channels}
\end{center}
\end{figure}

\begin{figure}[t!]
\begin{center}
\includegraphics[width=\qq]{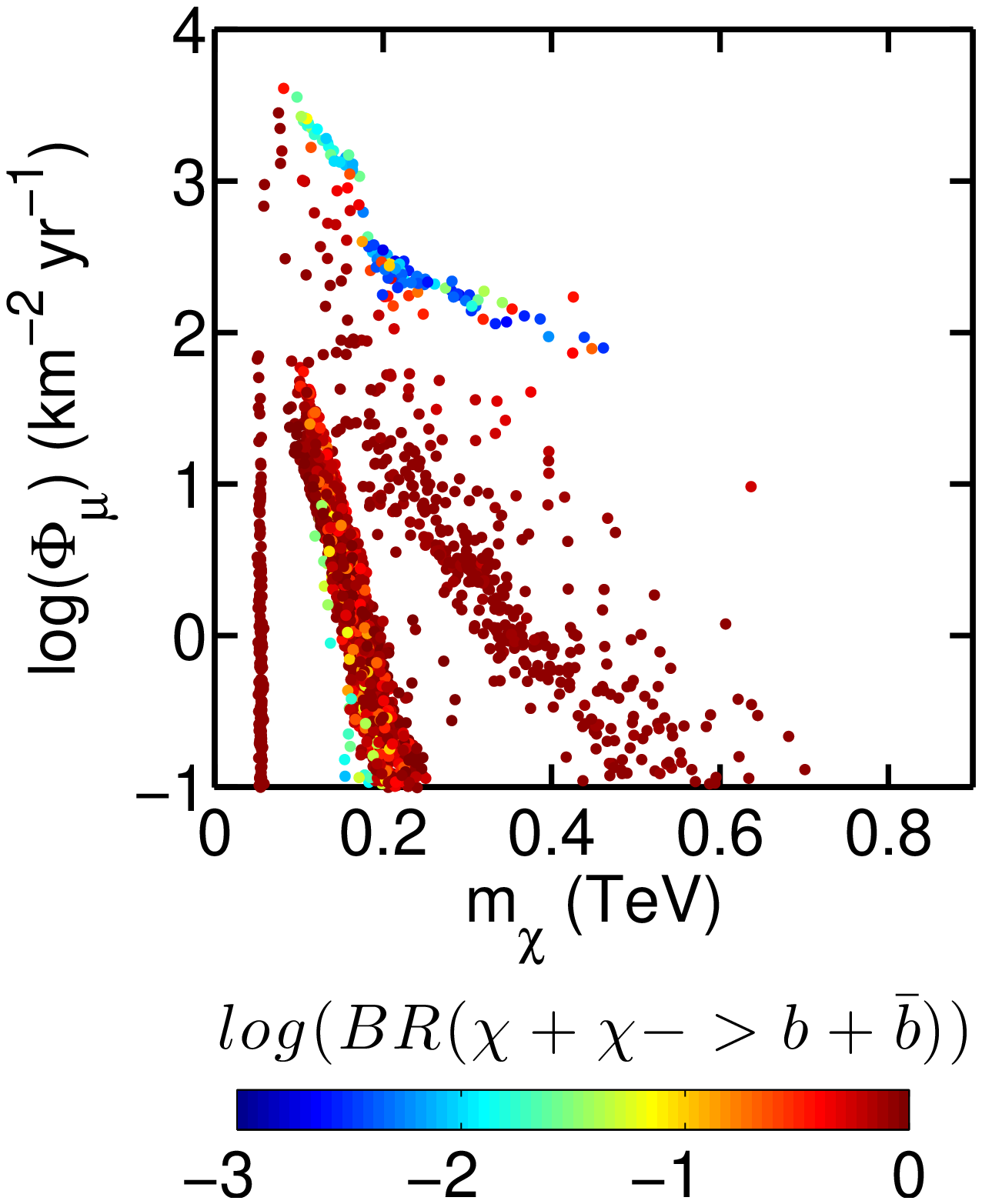}\hfill
\includegraphics[width=\qq]{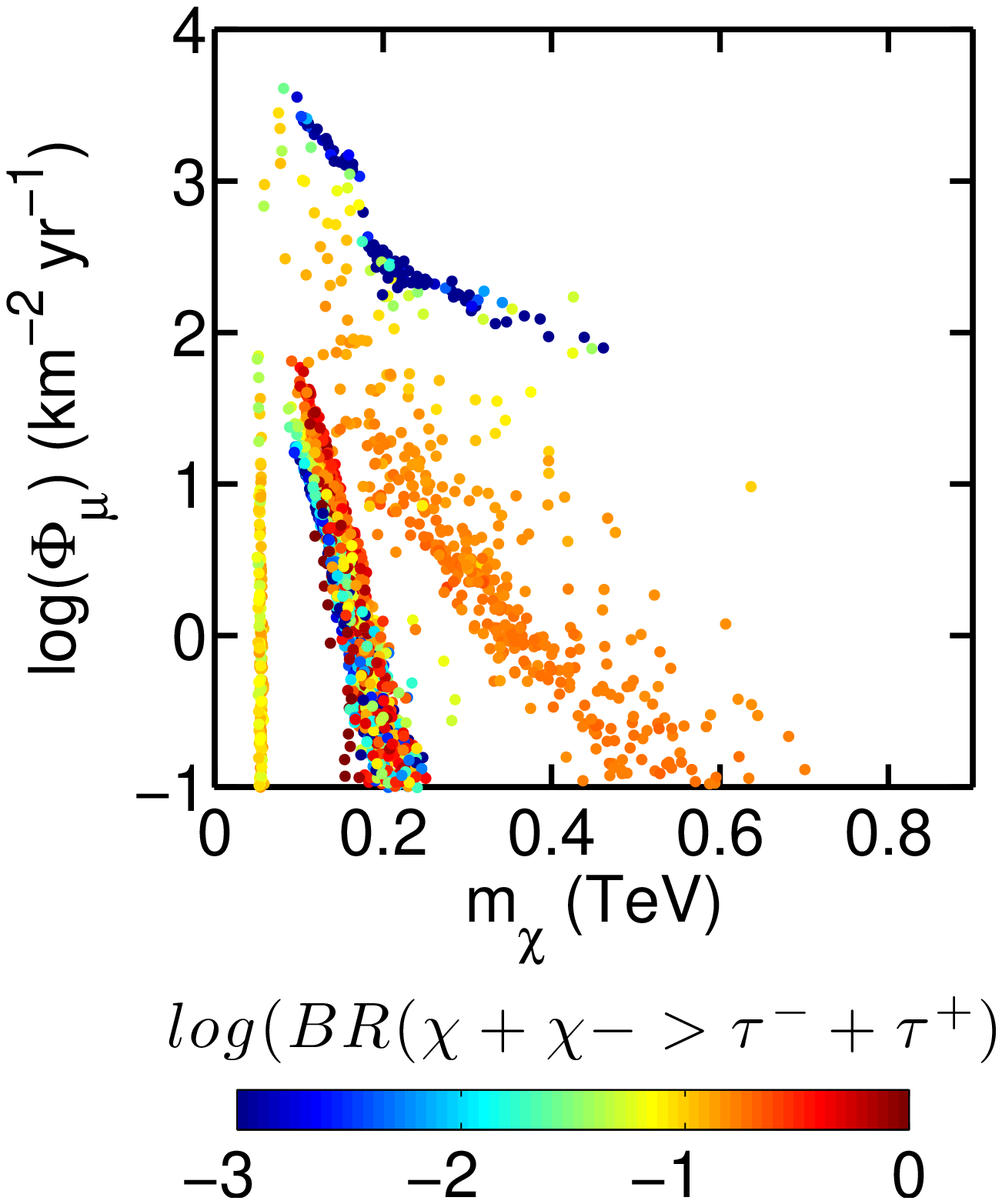}\hfill
\includegraphics[width=\qq]{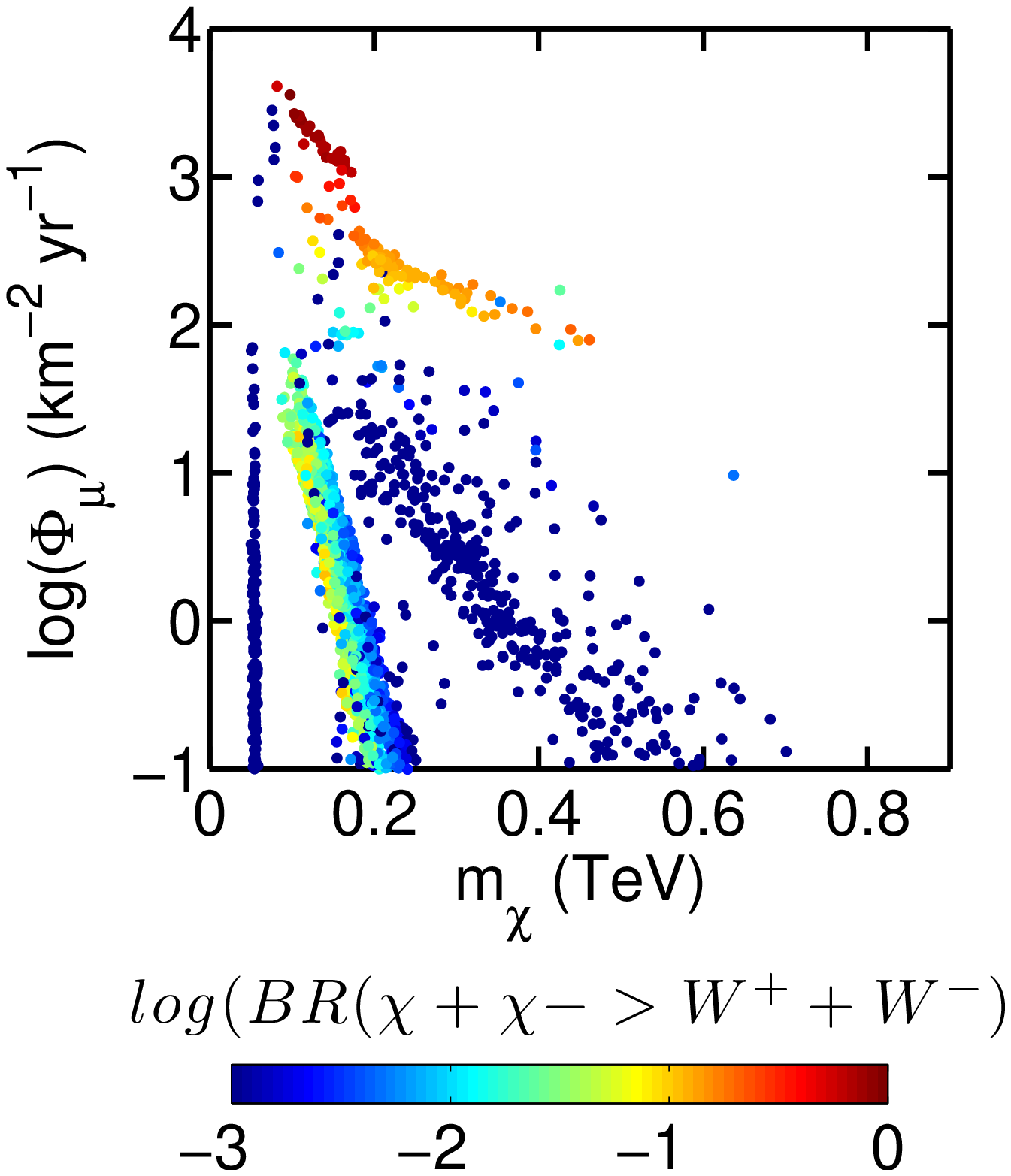}\hfill
\includegraphics[width=\qq]{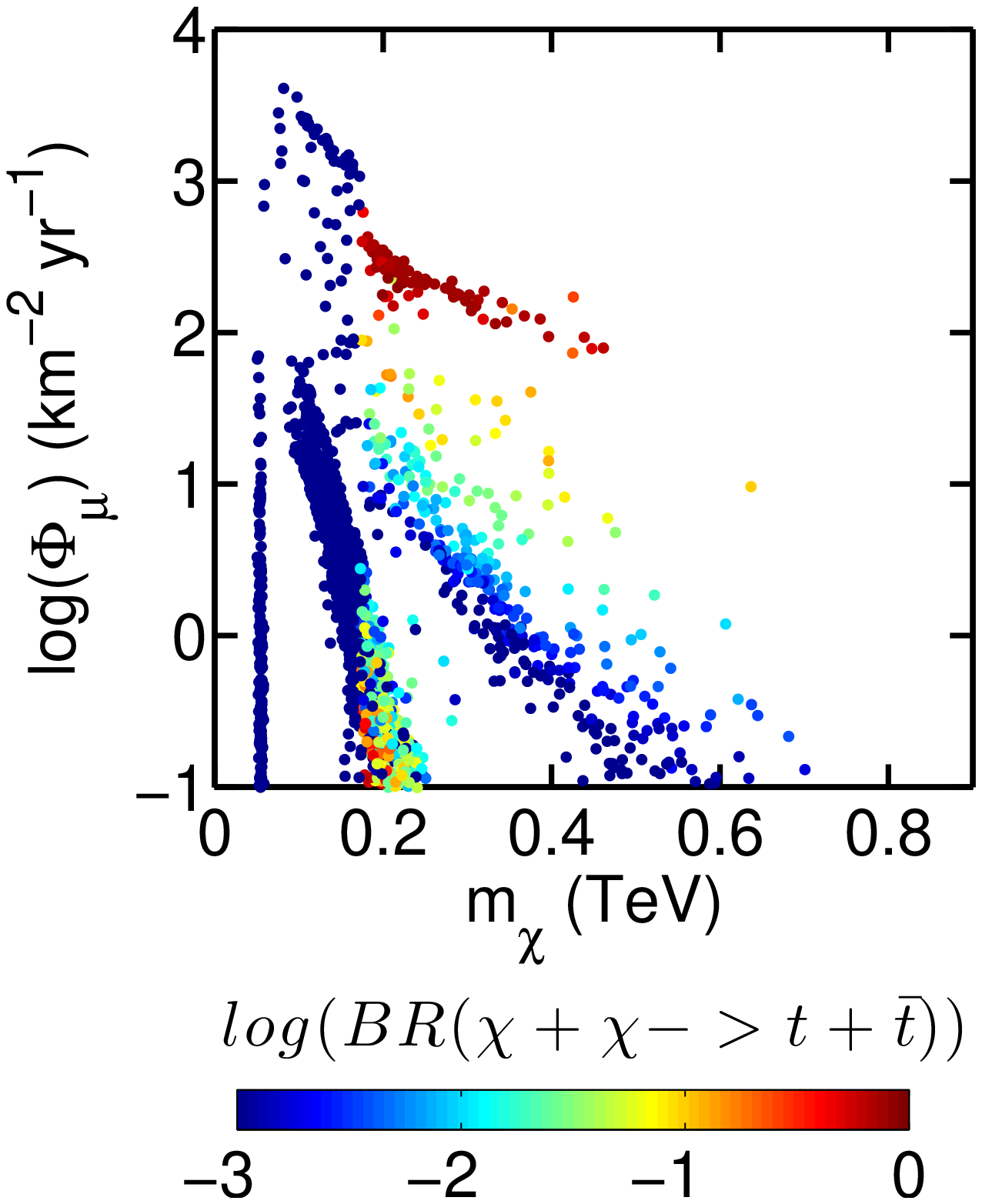} \\
\includegraphics[width=\qq]{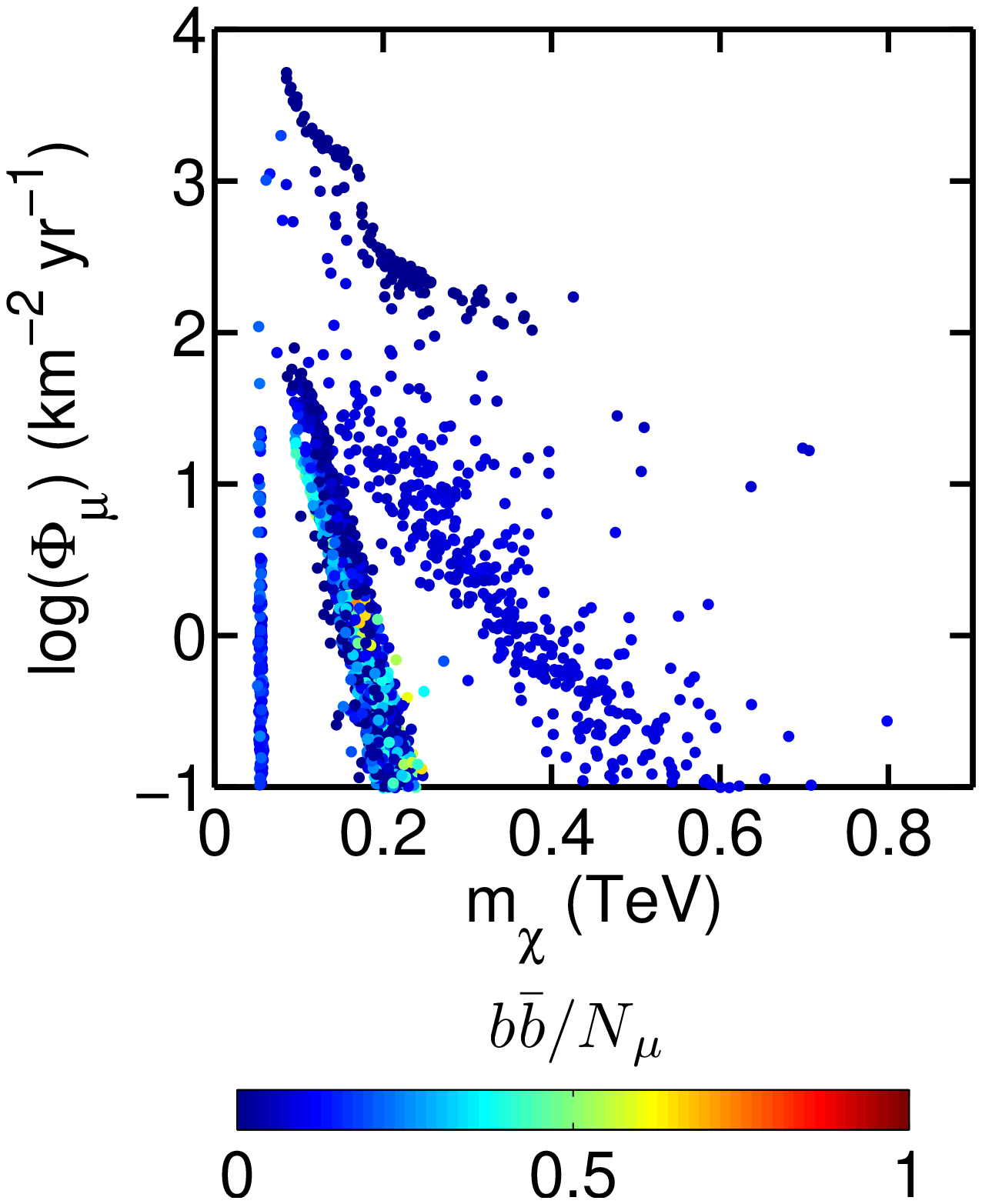}\hfill
\includegraphics[width=\qq]{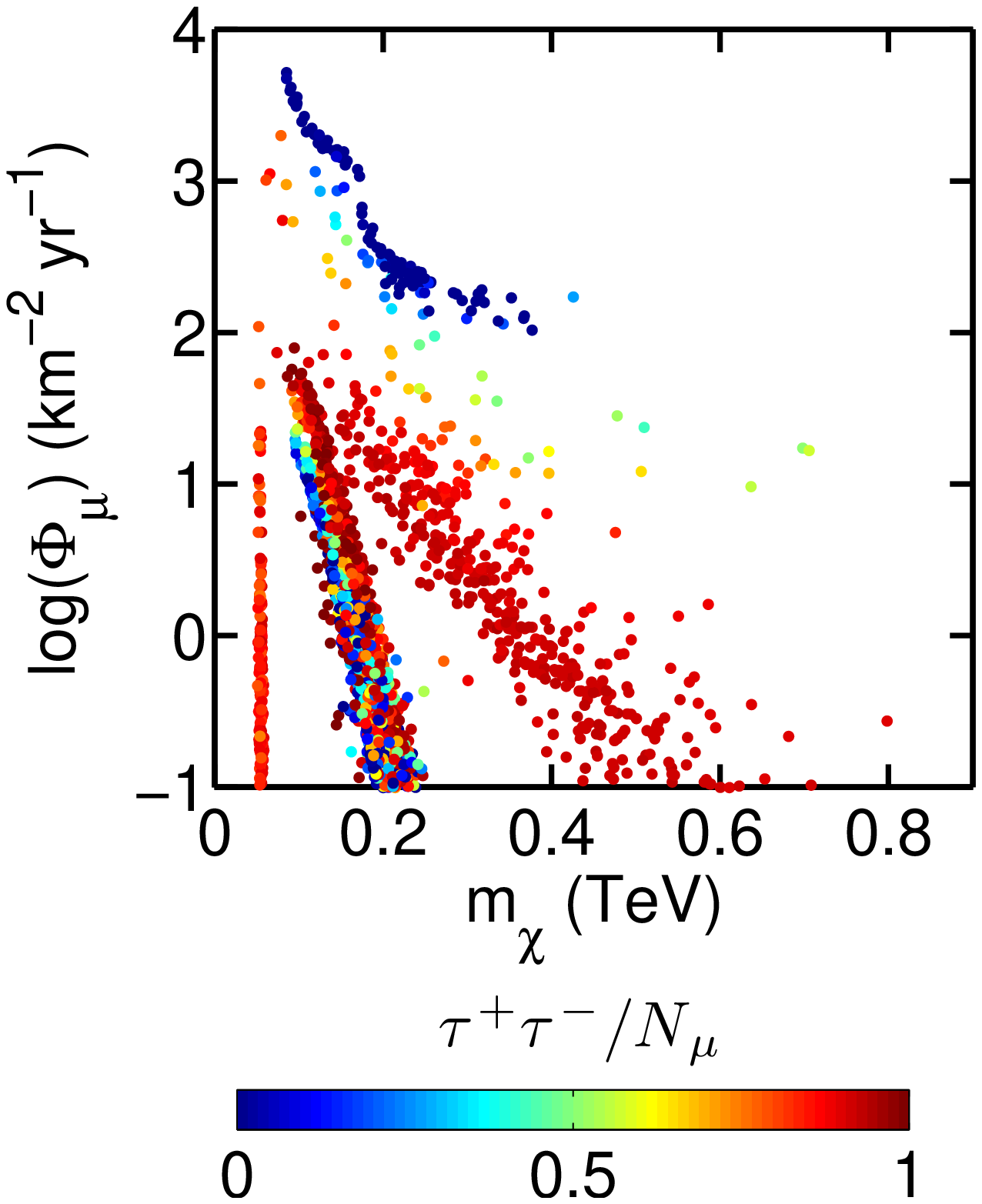}\hfill
\includegraphics[width=\qq]{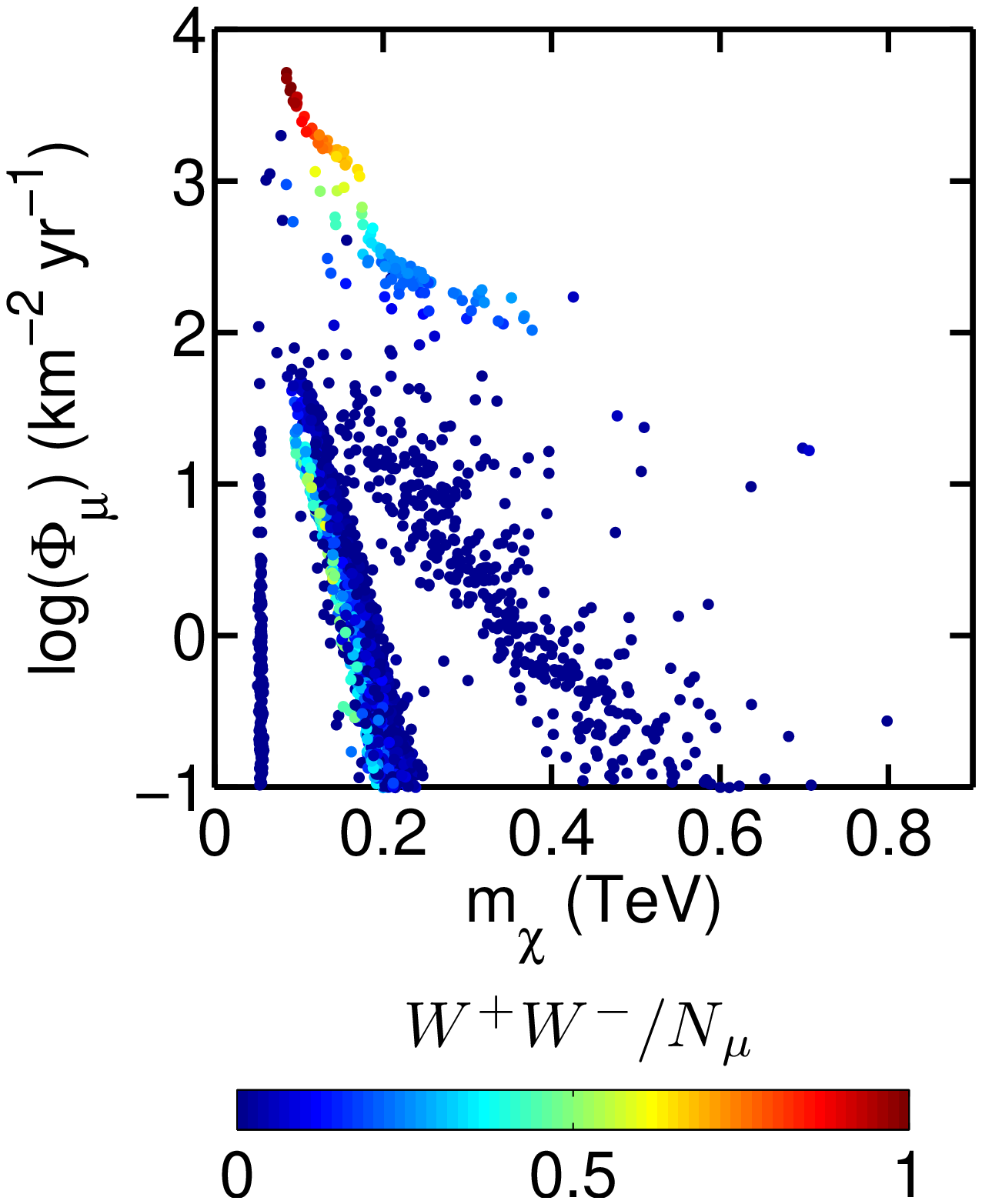}\hfill
\includegraphics[width=\qq]{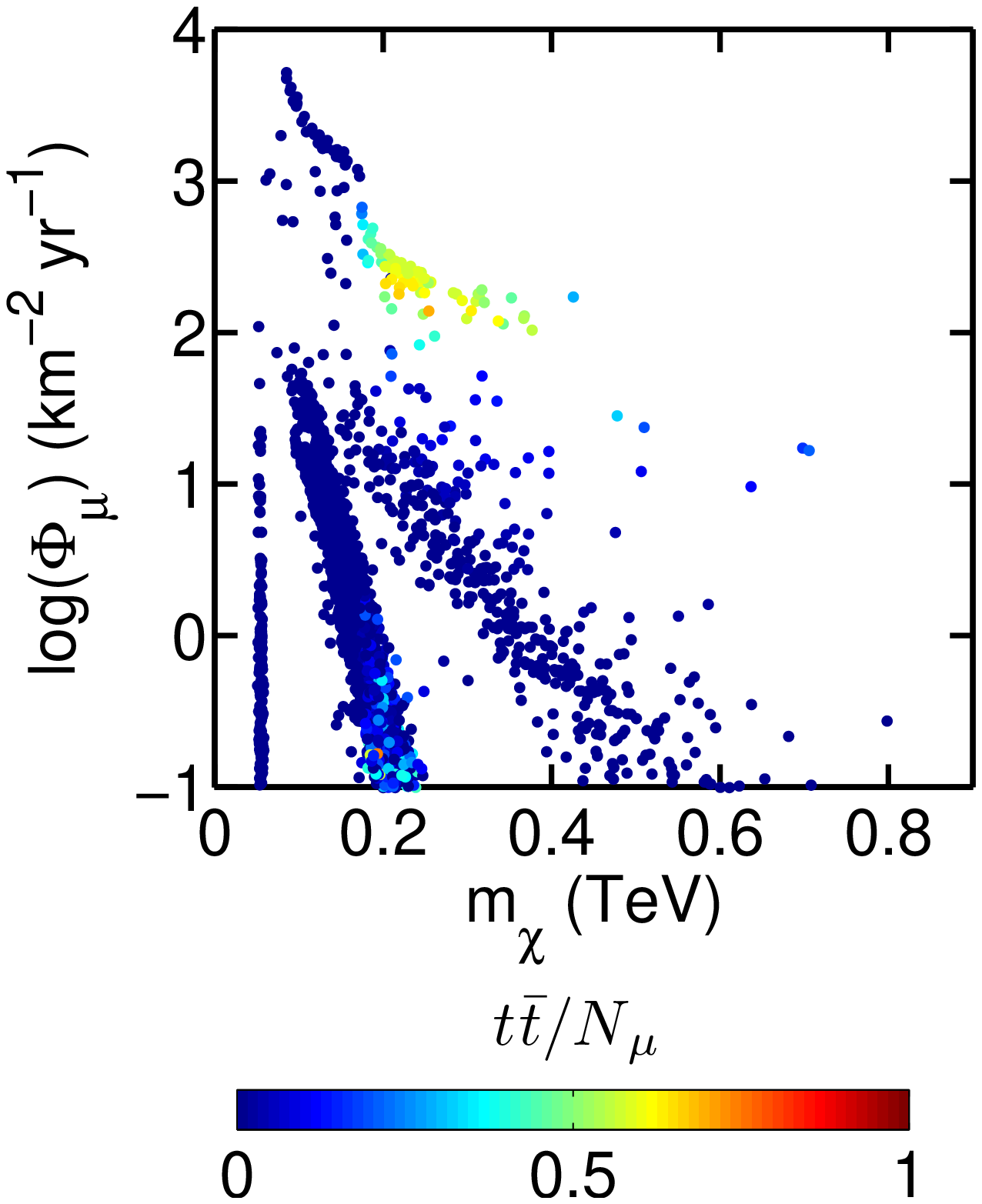}
\caption[test]{As in Fig.~\ref{fig:channels} but in the ($m_\chi, \Phi_\mu$) plane instead, where $\Phi_\mu$ is the muon flux 
above 1 GeV.}
\label{fig:channels_Phimu_mchi}
\end{center}
\end{figure}

We now turn to investigating the relative importance of the various final states across the whole CMSSM parameter space favoured by 
current data. The most important annihilation channels are summarized in Table~\ref{tab:channels}, while in Fig.~\ref{fig:channels} 
we plot the branching ratios (top row) and the fractional number of muon events (bottom row) in the $(\mhalf, \mzero)$ plane for 4 out 
of the 5 annihilation channels considered in the table (the plots for $ZZ$ are qualitatively similar to the ones involving $W^+W^-$ 
and are not shown). The same quantities but in the $(m_\chi, \Phi_\mu)$ plane are shown in Fig.~\ref{fig:channels_Phimu_mchi}. Although 
the dominant annihilation mode is $\chi \chi \rightarrow b\bar{b}$, whose average branching ratio is $0.67$, this soft channel actually 
only contributes an average 4\% to the total number of events. As argued above, this is because most of the differential muon flux 
from $\bbar$ is at low energies ($\lsim 10 \gev$, see Fig.~\ref{fig:diff_flux}), below the threshold energy adopted for IceCube. 
Therefore after convolution with $\Aeff_\mu$ the resulting number of muon events is a small fraction of the total. Clearly, a lowering 
of the threshold energy and an increase of the effective area at small muon energies would allow the DeepCore configuration to increase 
significantly its sensitivity to this soft channel. The second most important channel (in terms of its average branching ratio) is 
$\chi \chi \rightarrow \tau^+\tau^-$, which leads to a harder spectrum and therefore to a larger fraction of the number of events after 
convolution with the effective area. Indeed, the $\tau^+\tau^-$ is responsible for $\sim 65\%$ of the total number of events.  By 
comparing the plot of the fractional number of events in the $\tau^+\tau^-$ channel (second panel from the left in the bottom row of 
Fig.~\ref{fig:channels}) with the right panel of Fig.~\ref{fig:events_and_sigma_3D}, it is clear that this annihilation mode dominates 
in a region of the CMSSM parameter space which will remain undetectable by IceCube. On the other hand, if we consider 
the parameter space region within the reach of IceCube, we see from Fig.~\ref{fig:channels} that the most relevant channels will be the 
ones with gauge bosons final states ($W^+W^-$ and, not shown, $ZZ$), as well as the $\tbar$ final state, for the reasons explained 
above. In the parameter space accessible to IceCube, the two final states involving gauge bosons account together for about 80\% of 
the signal on average, while the $\tbar$ channel accounts for an average $14\%$, see Table~\ref{tab:channels}.

\begin{table}[t]
\centering
\begin{tabular}{| l | l l | l l |}
\hline
Channel $i$ &  $\langle \text{BR} \rangle_\text{CMSSM}$ &   $\langle N_i/N_\mu \rangle_\text{CMSSM}$
& $\langle \text{BR} \rangle_\text{IceCube}$ &   $\langle N_i/N_\mu \rangle_\text{IceCube}$ \\\hline
$\chi \chi \rightarrow b\bar{b}$ &  $0.67$     & $0.04$  & $6.6\times 10^{-2}$ & $6.7\times 10^{-4}$ \\     
$\chi \chi \rightarrow \tau^+\tau^-$ & $0.13$  & $0.65$  & $8.6\times 10^{-3}$ & $3.0\times 10^{-2}$ \\
$\chi \chi \rightarrow t\bar{t}$ &  0.07    & 0.08  & $0.26$ & $0.14$  \\
$\chi \chi \rightarrow W^+W^-$ &  $0.02$  & $0.11$ & $0.49$ & $0.60$\\
$\chi \chi \rightarrow ZZ$ & $5\times10^{-3}$  & $0.03$ & $0.14$ & $0.20$  \\\hline
\end{tabular}
\caption{Dominant annihilation channels and the associated fractional number of events, averaged over the posterior of the CMSSM 
parameter space (columns 2 and 3) and averaged over the parameter space accessible to IceCube (defined as the region with S/N $\geq 5$, 
columns 4 and 5). \label{tab:channels}}
\end{table}

In summary, in the context of the CMSSM the overall dominating annihilation mode in terms of its branching ratio 
is $\chi \chi \rightarrow b\bar{b}$, although we have found that this contributes only a very small fraction to the number of 
events that are within reach of IceCube. In the parameter space region accessible to IceCube, though, the most important annihilation 
channels (which dominate the number of events) are the ones involving gauge bosons final states as well as $\chi \chi \rightarrow \tbar$.
  We have also found that the least biased choice for a single--channel reconstruction is $\chi \chi \rightarrow W^+W^-$, which 
together with $\chi \chi \rightarrow ZZ$ (which has a very similar spectral shape) accounts for $\sim 80\%$ of the number of events 
in the CMSSM parameter space accessible to IceCube.

\section{Comparison with direct detection techniques} \label{sec:DD}

\subsection{Ruling out the CMSSM with IceCube and direct detection}

It is interesting to consider the cross--correlation between the signal in a neutrino telescope and the expected signal from direct 
detection methods, as this gives a consistency check of the model. 
To this aim, we computed the neutralino spin--independent scattering cross section off a proton according to the prescription given in Ref.~\cite{Drees:1993bu}. When comparing the indirect detection prospects using neutrino telescopes with direct detection experiments, it is important to keep in mind that predictions for the latter also suffer from a number of sizable uncertainties. The value of the neutralino-nucleon elastic scattering cross section is subject to up to a factor of $\sim 3$ ($\sim 10$) uncertainty at 68\% C.L. (at 95\% C.L.) due to the estimate of the hadronic matrix elements~\cite{Hadronic_uncertainties}. On the other hand, uncertainties in the form factors appear to be subdominant~\cite{Green:2007rb}. When comparing experimental counts with theoretical predictions, the local WIMP distribution has to be taken into account, and in particular the local WIMP density $\rholoc$ and their velocity distribution. Although it has generically been assumed that $\rholoc$ suffers from uncertainties of up to a factor of 2 or 3, recent studies have demonstrated that this crucial quantity can be determined to 10\%--20\% accuracy by using a variety of kinematic data to constrain a parameterized model of the galactic halo~\cite{MW_DD}. Furthermore, given at least two experiments with different target nuclei masses, the WIMP mass and scattering cross section can be reconstructed independently of the velocity distribution of the halo~\cite{Indep_VV}. It is presently difficult to quantify the potential impact of other astrophysical sources of uncertainties, such as for example a non-smooth component to the local density and the effect of a co-rotating dark matter halo. For definiteness, when showing below direct detection constraints on the spin-independent cross section from direct detection experiments, we adhere to the standard convention of assuming a smooth halo with $\rholoc = 0.3$ GeV/cm$^3$ and a Maxwellian velocity distribution with mean velocity $\bar{v} = 220$ km/s.

\begin{figure}[t!]
\begin{center}
\includegraphics[width=\ww]{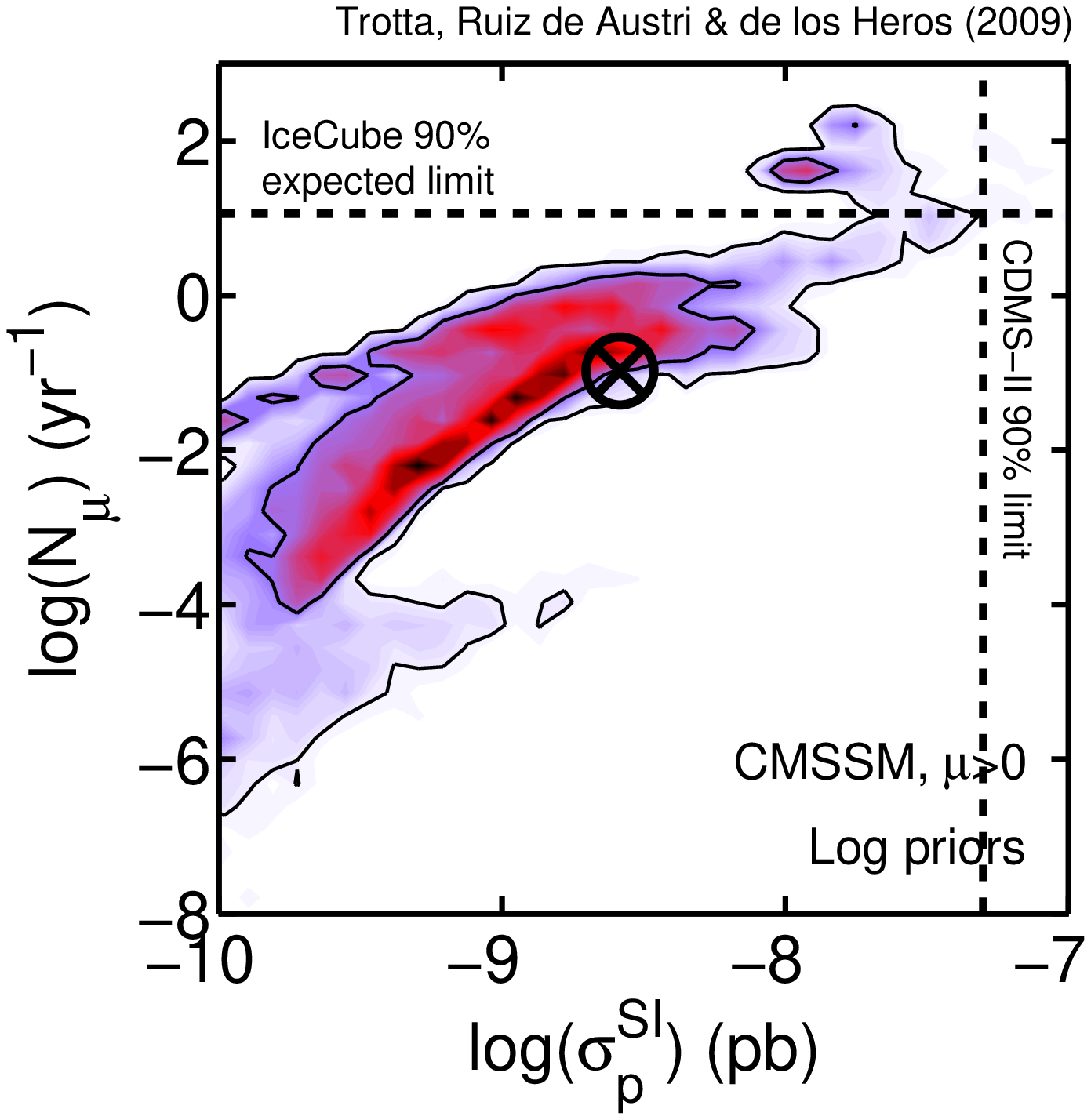}
\includegraphics[width=\ww]{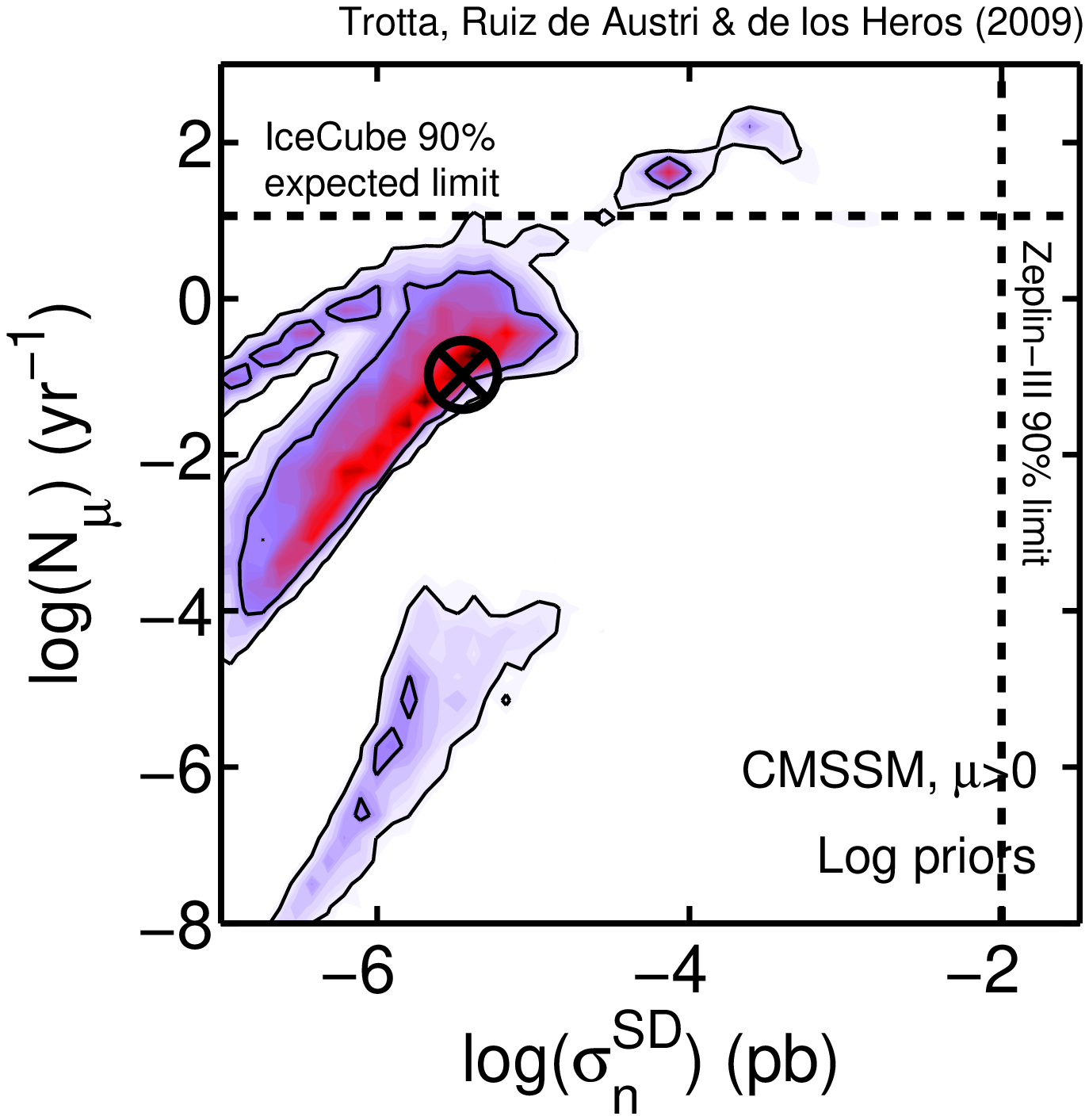}
\caption[test]{2D probability distribution for the number of muon events in IceCube vs the spin--independent (left panel) and 
spin--dependent (right panel) scattering cross section (the theoretical predictions are very similar for scattering off a proton or 
off a neutron). The vertical lines give the current best 90\% upper limits from direct detection experiments~\cite{cdms08si,zeplin-IIIsd}
  (for $\mchi = 100 \gev$), the horizontal line delimits the parameter space that IceCube will be able to exclude with 90\% confidence.} \label{fig:events_vs_DD}
\end{center}
\end{figure}

Figure \ref{fig:events_vs_DD} shows the 68\% and 95\% favoured 
regions in the plane spanned by the number of muon events expected for IceCube  vs the neutralino scattering cross section off a 
nucleon (spin--independent scattering off a proton, $\sigsip$, in the left panel and spin--dependent scattering off a neutron, $\sigsdn$, 
in the right panel). As expected, the region of parameter space leading to a large number of events for IceCube also predicts larger 
scattering cross sections. Current 90\% upper limits for the spin--independent cross section are around $\sigsip \lsim 5 \times10^{-8}$
~pb~\cite{cdms08si,zeplin-IIIsi,xenon-10si} for a 100 GeV neutralino, while the spin--dependent cross sections are constrained by 
direct detection experiment $\sigsdp \lsim 2 \times10^{-1}$~pb and  $\sigsdn \lsim 10^{-2}$~pb, for $\mchi \sim 100 \gev$
~\cite{zeplin-IIIsd,xenon-10sd,coupp-08sd,kims07sd,naiad05sd}. The SuperKamiokande neutrino detector puts more stringent constraints 
on the spin dependent scattering cross section off a proton, with an upper limit $\sigsdp \lsim 4 \times10^{-3}$~pb~\cite{superk04sd}, while the current IceCube data provide an upper limit $\sigsdp \lsim 3 \times10^{-4}$~pb (although this limit applies for a slightly higher neutralino mass of 250 GeV)~\cite{IceCube:09a}.

It follows from the left panel of Fig.~\ref{fig:events_vs_DD} that should IceCube detect neutrinos from the Sun at high significance, 
then the corresponding spin--independent cross section must satisfy $\sigsip \gsim 0.9 \times 10^{-8}$ pb. Therefore, another factor 
of $\sim 5$ improvement in the direct detection constraints on the spin--independent scattering cross section would rule out the CMSSM 
parameter space region in which IceCube would be able to detect a signal. On the other hand, if IceCube does detect neutrinos from the 
Sun at high significance while direct detection experiments fail to find a signal above $\sigsip \gsim 0.9 \times 10^{-8}$ pb, this 
would strongly disfavour the CMSSM, at least given the conventional astrophysical assumptions made in the present analysis.
Changes in the local WIMP density would equally affect the signal for both direct detection and neutrino telescopes, at least for points satisfying the equilibrium condition, so the relative 
power would remain unchanged (although of course the absolute prospects of detection would be modified). The same applies with respect to the uncertainties in computing the neutralino-nucleon scattering cross section. In the presence of a dark 
disk, the parameter space region accessible to IceCube increases as discussed above, but it still corresponds to values of the 
spin--independent scattering cross section $\sigsip \gsim 0.7 \times 10^{-8}$ pb. We have found that the above results are robust within 20\% under a change of priors for the CMSSM parameters, as discussed in section~\ref{sec:prior_change}.

\subsection{The capture--annihilation equilibrium condition}
\label{sec:equilibrium}

\begin{figure}[tbh!]
\begin{center}
\includegraphics[width=\ww]{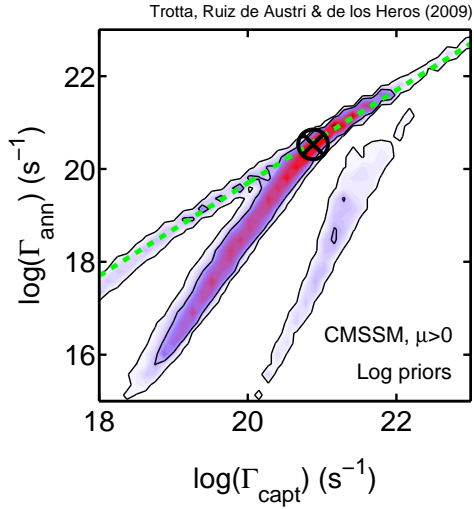} 
\caption[test]{2D probability distribution for the annihilation rate in the center of the Sun, $\GA$, vs the capture rate of WIMPs in 
the Sun, $\Gcap$, in the CMSSM. The diagonal, dashed green line denotes the equilibrium condition, where capture and annihilation are 
balanced.}
\label{fig:rates_2D}
\end{center}
\end{figure}

\begin{figure}[tbh!]
\begin{center}
\includegraphics[width=\ww]{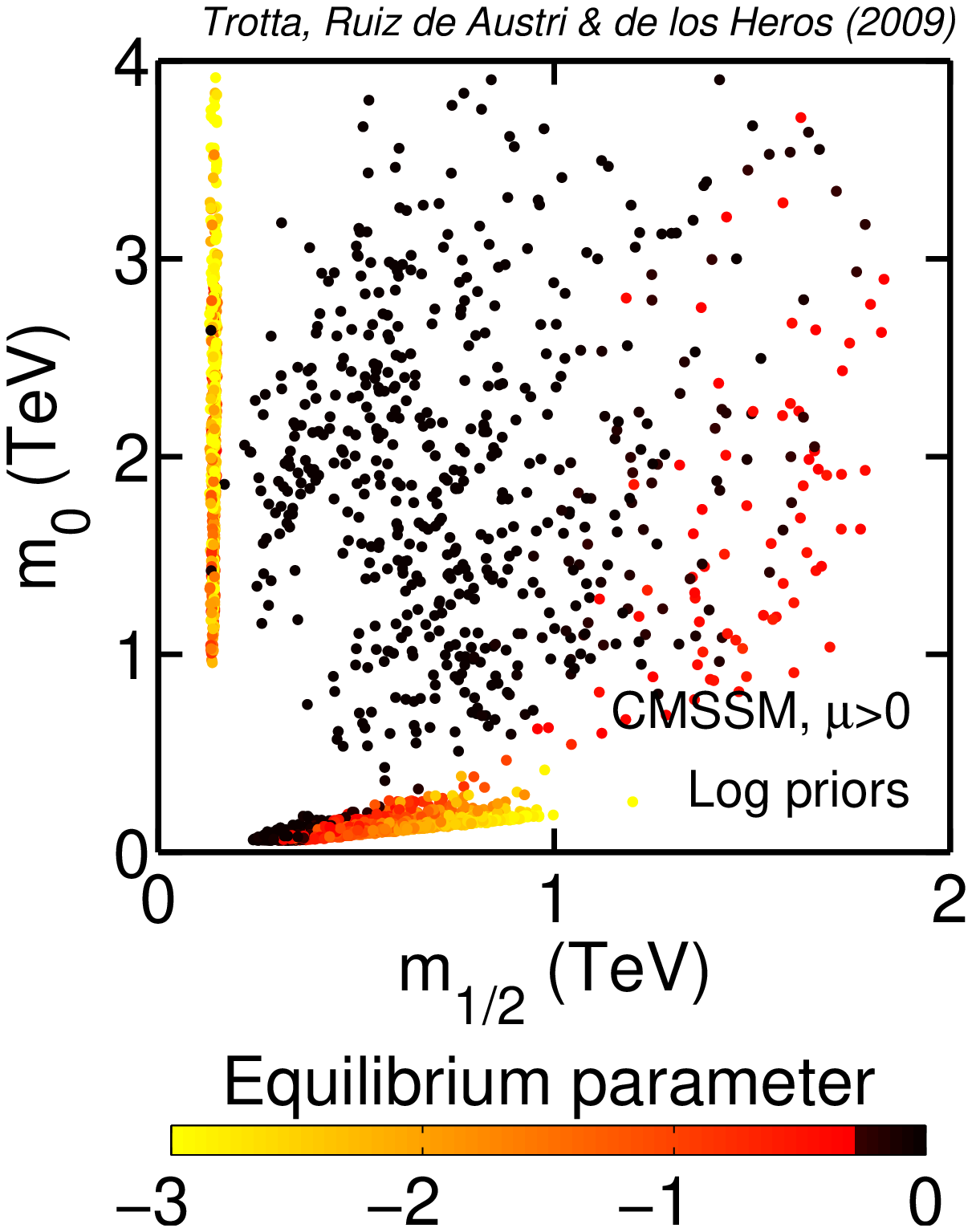}
\caption[test]{The distribution of the equilibrium parameter $\Ep$ defined in Eq.~\eqref{eq:defep} (colour coding) in the 
($\mhalf, \mzero$) plane. Regions in parameter space with $\Ep > -0.3$ (dark points) have reached equilibrium between capture and 
annihilation in the center of the Sun. Although about 66\% of the parameter violates the equilibrium condition, the focus point region 
accessible to IceCube is in equilibrium.}
\label{fig:equilibrium_3D} 
\end{center}
\end{figure}

In comparing the constraints obtained from searches using neutrino telescopes with the limits obtained via direct detection techniques, 
it is often assumed that equilibrium is reached between the annihilation and capture rates in the Sun, (see e.g.,~\cite{IceCube:09a}, 
and for a recent discussion~\cite{Wikstrom:2009kw}). This means from Eq.~\eqref{eq:GA} that  $\GA= \Gcap/2$, and therefore limits 
on $\GA$ from neutrino telescopes can be directly translated into corresponding limits for the scattering cross sections 
(either spin--dependent or spin--independent), which are proportional to $\Gcap$. It is thus important to verify whether the equilibrium 
condition is indeed achieved across the CMSSM parameter space. We plot the 2D probability distribution for $\GA$ and $\Gcap$ in 
Fig.~\ref{fig:rates_2D}.  The region of the CMSSM parameter space that indeed has reached equilibrium corresponds to the focus point 
(around the dashed, diagonal green line, in the region with larger values of $\GA$), and to parts of the stau coannihilation region 
(cf.~Fig.~\ref{fig:equilibrium_3D},
 showing the  equilibrium condition in the ($\mhalf, \mzero$) plane). However, we can see that in other parts of the coannihilation 
region (branching off diagonally from the best fit point, denoted by an encircled cross) and the $h$--pole region (isolated diagonal 
island at 95\% probability) the annihilation rate is very significantly lower than the capture rate, and hence equilibrium is not 
attained. This also explains the low number of muon events expected in those parts of parameter space. 

In order to quantify the extent to which the CMSSM parameter space is in equilibrium, we define the equilibrium parameter
\be \label{eq:defep}
\Ep \equiv \log\frac{2\GA}{\Gcap},
\ee
where $\Ep = 0$ denotes equilibrium. We consider that equilibrium is not reached if $\GA$ is below a factor of 2 the equilibrium 
condition, i.e.\ for $\Ep < -0.3$. The value of the equilibrium parameter in the ($\mhalf, \mzero$) plane is plotted in 
Fig.~\ref{fig:equilibrium_3D}, confirming that while points in the focus point have reached equilibrium, most samples in the 
coannhiliation region and in the $h$-pole region are out of equilibrium. Indeed, $\sim 66\%$ of the samples across the parameter 
space violate the equilibrium condition. However, the region accessible to IceCube is delimited by $\Gcap \gsim 10^{22}~\text{s}^{-1}$ 
and $\GA \gsim 5\times10^{21}~\text{s}^{-1}$, and this part of parameter space is indeed in equilibrium, as would be expected.

\section{Prior dependency}
\label{sec:prior_change}

\begin{figure}[t!]
\begin{center}
\includegraphics[width=\ww]{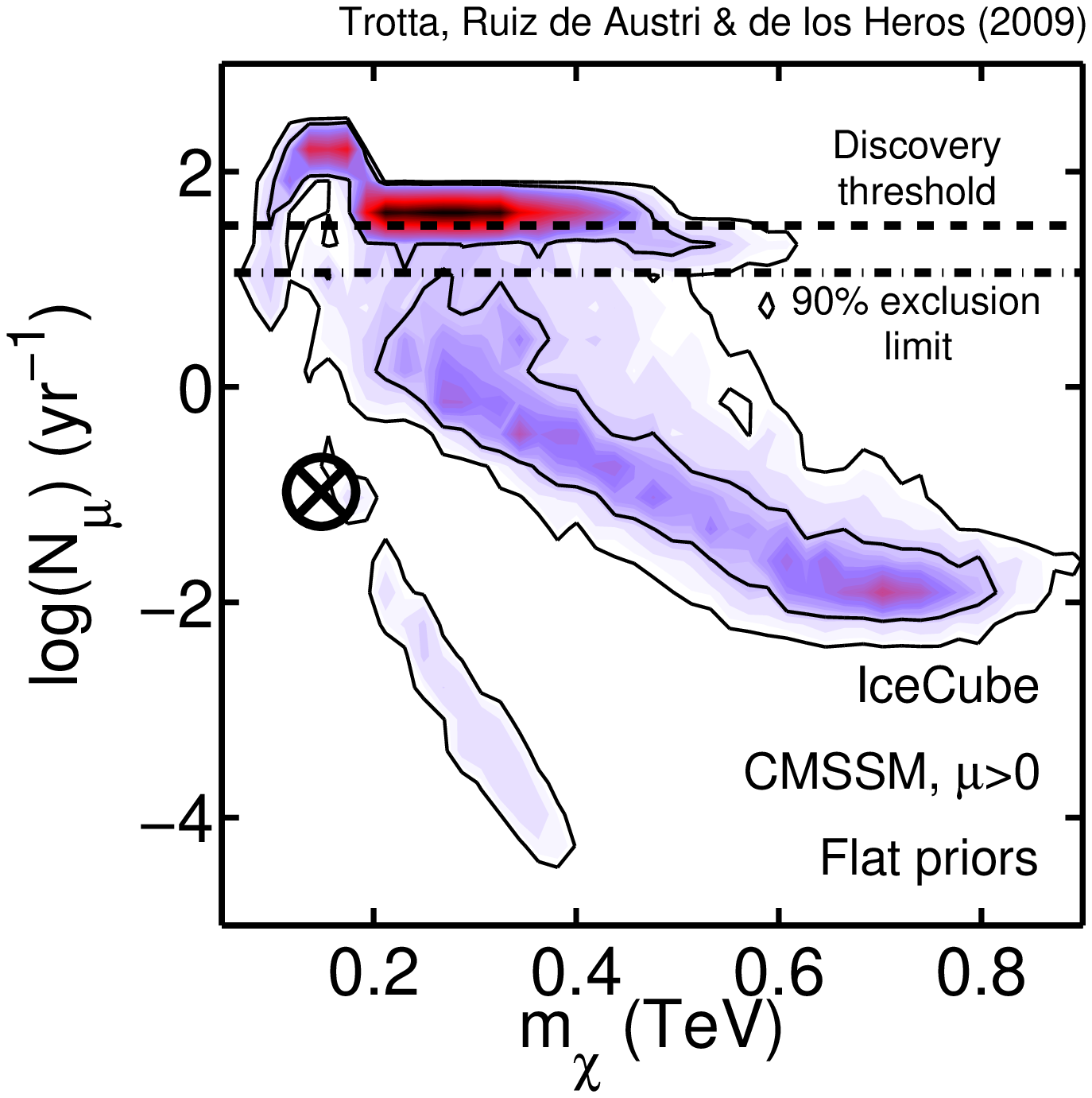}
\includegraphics[width=\ww]{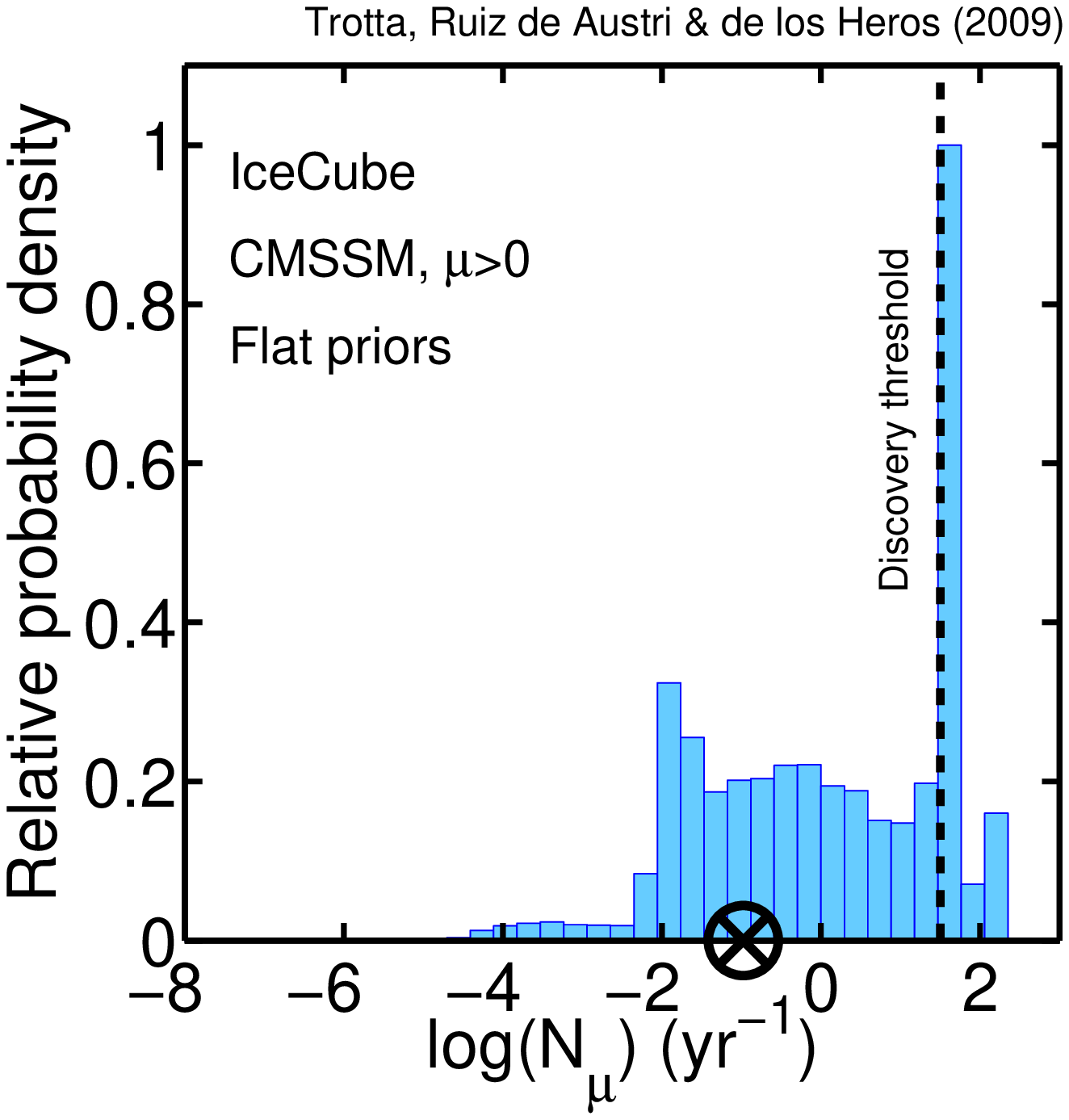}
\caption[test]{Left panel: probability distribution for the number of muon events vs neutralino mass, adopting a flat prior on the 
CMSSM parameters. Right panel: marginalized 1D probability distribution for the number of muon events. The black encircled cross is 
the best fit (cf.~Fig.~\ref{fig:muevents}, for a log prior). }\label{fig:events_flat_prior}
\end{center}
\end{figure}

The dominant source of uncertainty in our prediction is the choice of priors for the CMSSM scalar and gaugino masses. All results 
presented until here employed a log prior, but other choices are possible (for more details, see 
e.g.~\cite{Allanach:2008iq,tfhrr:2008,bclw07}). 
Since the prior choice sets the probabilistic measure of for the parameter space, it has been shown that certain classes of priors can naturally quantify the physical fine-tuning of the theory~\cite{allanach06,ccr08}.
One however expects the choice of prior to become irrelevant 
once the data included in the likelihood are strongly constraining, a limit not yet attained by present--day data but which will be obtained by future LHC measurements~\cite{LHC:inprep}. The main effect of 
a change of priors is to modify the relative importance of the focus point region with respect to the coannihilation region. Rather than performing a complete exploration of several classes of priors, here we simply investigate the robustness of our results with respect to a change of prior on the CMSSM masses by 
showing in Fig.~\ref{fig:events_flat_prior} results obtained assuming a flat prior on ($\mzero, \mhalf$) over the range 
$50 \gev \leq \mzero, \mhalf \leq 4 \tev$, with all other aspects of the scan unchanged (compare with Fig.~\ref{fig:muevents}, 
adopting a log prior). Such a choice is expected to roughly bracket the results one would obtain when analysis a larger class of priors. We can see that under a flat prior the preference for the low mass coannhilation region is attenuated, and 
there is a much larger statistical weight on the focus point region (which will be accessible to IceCube). This comes about because 
under the flat prior, the volume encompassed by the mass range between e.g. 1~TeV and 4~TeV is a factor $100$ larger than between 
100~GeV and 400~GeV, while under the log prior the ratio of the volumes of the two regions is unity. Since the preference for 
the low--mass region encoded in the likelihood is not sufficiently strong to override the much larger prior volume of the focus 
point under the flat prior, the ensuing constraints remain at this point somewhat prior dependent. 

Using the flat prior, we find that the parameter space region accessible to IceCube at the $5\sigma$ level is larger than what we had 
found under the log prior, and is now about 12.5\% of the total parameter space. An even more pronounced boost of the detection 
prospects is obtained if one introduces theoretical priors that further favour the focus point region, as has been done in 
Ref.~\cite{Allanach:2008iq}. A detailed comparison with the rather more optimistic detection prospects for IceCube claimed in 
Ref.~\cite{Allanach:2008iq} is difficult, as there can be several differences in our analyses.  In Ref.~\cite{Allanach:2008iq} not many
details are given about their treatment of neutrino oscillations, nor about the effective area employed for the telescope. In our 
case, we have used a state--of--the--art code for our prediction of the neutrino flux at the Earth, which fully 
accounts for neutrino oscillations, and convolved the flux with the energy--dependent effective area obtained by Monte Carlo studies 
of the telescope. Although their ``flat $\tanb$ prior'' corresponds to our flat prior analysis, they seem to obtain a sizable 
probability of several thousands of events per year (compare the dashed line in the right panel of Fig.~8 in Ref.~\cite{Allanach:2008iq} 
with the right panel of our Fig.~\ref{fig:events_flat_prior}).

\begin{figure}[t!]
\begin{center}
\includegraphics[width=\ww]{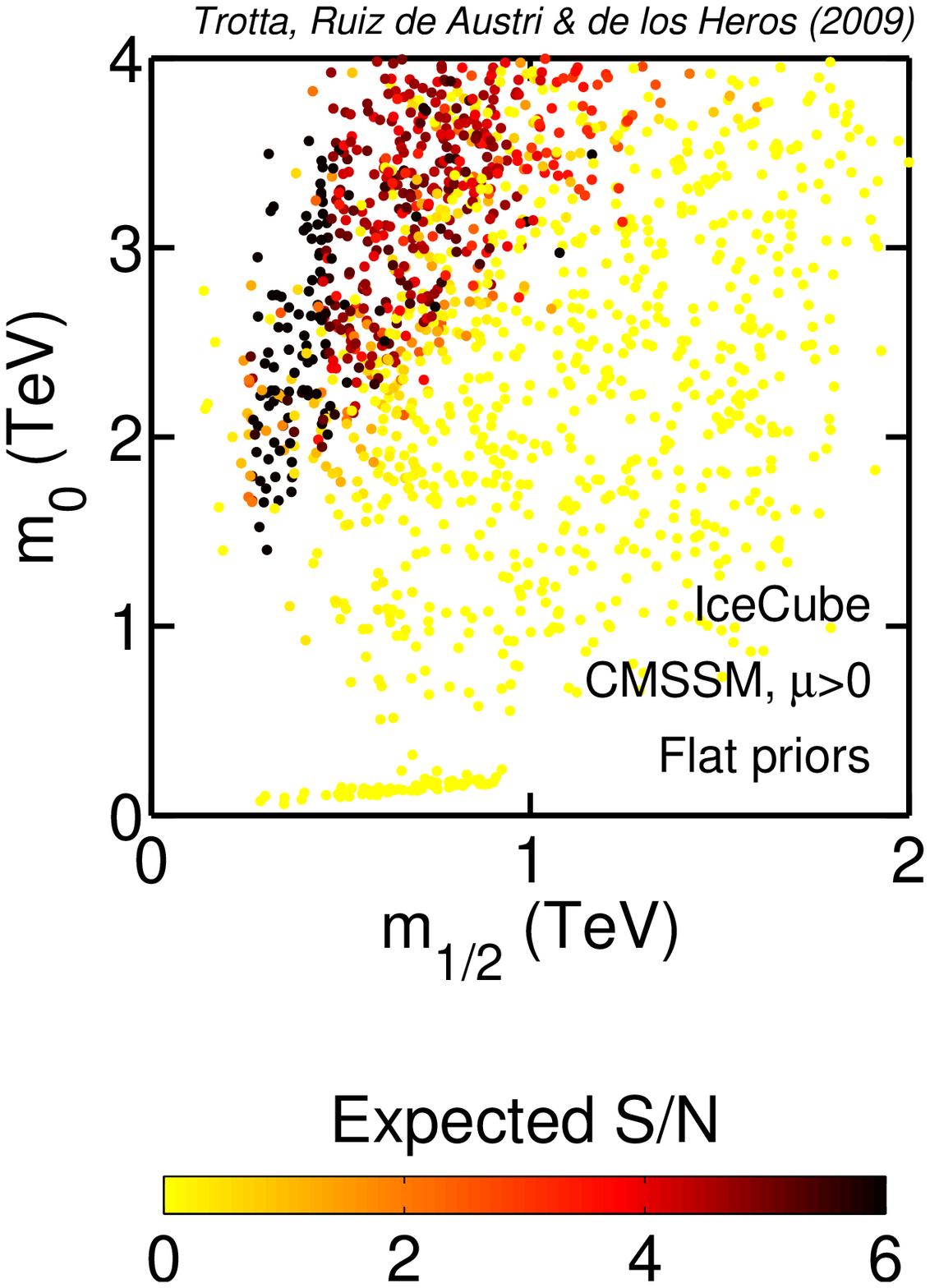}
\caption[test]{Signal--to--noise levels over the background for the WIMP signal from the Sun (cf.~with Fig.~\ref{fig:events_and_sigma_3D}, which adopts a 
log prior on the CMSSM masses instead).}\label{fig:events_and_sigma_3D_flat_prior}
\end{center}
\end{figure}

 The authors in  Ref.~\cite{Bruch:2009rp} have studied the effect of a possible dark matter disk in the galaxy on the capture of neutralinos 
in the Sun and the corresponding expected muon flux at the Earth. They, however, do not make concrete predictions on the number of 
expected events in any particular detector, as we do in this work, but just compare the predicted flux with existing experimental limits and with projected IceCube limits, both derived using the assumptions that a hard channel dominates the spectrum (which we have shown can introduce systematic errors). 
If we, very simplistically, mimic the effect of such disk by increasing our predictions by an order of magnitude (the effect found in Figure~1 in~\cite{Bruch:2009rp}) 
we obtain very similar results to Ref.~\cite{Bruch:2009rp} in terms of the flux expected in IceCube. Using flat priors and in the presence of a dark disk, the IceCube accessible region increases to about 42\% of the CMSSM 
parameter space. This boost in the probability of a discovery comes about because under the flat prior there is a large region of high probability density 
coming from the focus point that sits just below the IceCube sensitivity under normal astrophysical assumptions (see the left panel of Fig.~\ref{fig:events_flat_prior}), but when a dark disk is added to the picture, its larger capture rate decisively thrusts this region above the detection threshold. However, this scenario has to be considered 
as very optimistic, in that it relies both on astrophysical and statistical effects (coming from theoretical prior consideration) to increase by a large 
factor the probability of discovery. Furthermore, it would seem from the left panel of Figure 4 in Ref.~\cite{Bruch:2009rp} that such a large boost to the muon flux would be already under pressure from current neutrino telescope data, and therefore if the dark disk exists its properties remain at the moment quite uncertain.
In conclusion, it seems more reasonable to consider the $\sim 5\%$ probability of discovery obtained under the log prior and the 
dark disk assumption as a more realistic upper limit on the IceCube discovery potential.

\section{Conclusions}
\label{sec:conclusions}

We have studied in detail the ability of the nominal IceCube configuration to probe the favoured parameter space of the CMSSM, including 
a realistic description of the detector and the expected background from atmospheric neutrinos and the background from the solar 
corona.  We have however taken several conservative assumptions on the performance of IceCube in our calculations, so the results 
presented in this paper are conservative. 
The detection prospects of a WIMP induced signal from the Sun are best for IceCube in the focus point region, where equilibrium 
between capture and annihilation rate in the Sun is achieved and the signal is correspondingly boosted. We notice that this important 
part of parameter space will not be accessible to the LHC. Depending on the details of the analysis and on the astrophysical 
assumptions, we found that IceCube has between 2\% and 12\% probability of detecting a WIMP signal at more than $5\sigma$ significance. 

We have shown that the assumption that the signal is dominated by only one annihilation channel, as has become common 
practice in several experimental analyses, can lead to a considerable systematic bias in the conversion to the WIMP scattering 
cross section. The soft channels such as $\chi\chi \rightarrow \bbar$ are the most prone to this effect, while 
final states involving gauge bosons are the least affected. We therefore recommend adopting the $\chi\chi \rightarrow W^+W^-$ channel 
for the least bias single--channel reconstruction in the context of the CMSSM. 

By comparing the expected signal for IceCube with direct detection prospects, we have concluded that the CMSSM will be ruled out if 
IceCube does detect a WIMP signal from the Sun while direct detection experiments do not find anything above 
$\sigsip \gsim 0.7 \times 10^{-8}$~pb (for the more conservative choice of astrophysical model).  
We further find that increasing the sensitivity of IceCube to lower muon energies will boost its 
discovery potential of neutralino dark matter, since it will improve access to the $\chi \chi \rightarrow b\bar{b}$ annihilation 
channel, which is the dominating mode in the CMSSM. However this mode produces mainly low energy neutrinos, below the threshold of 
the original IceCube configuration. In this respect, the proposed low-energy DeepCore extension of IceCube will be an ideal 
 instrument to focus on the relevant CMSSM areas of neutralino parameter space.

\section*{Acknowledgements}
The authors would like to thank Tobias Bruch, Annika Peter and Timothy Sumner for useful conversations and suggestions. The work of R.R. has been supported in part by MEC (Spain) 
under grant FPA2007-60323, by Generalitat Valenciana under grant 
PROMETEO/2008/069 and by the Spanish Consolider-Ingenio 2010 Programme 
CPAN (CSD2007-00042). R.T. would like to thank the Galileo Galilei Institute for
Theoretical Physics for the hospitality and the INFN and the EU FP6 Marie Curie Research and Training Network ``UniverseNet''  
(MRTN-CT-2006-035863) for partial support. R.R. and R.T. would like to
thank the European Network of Theoretical Astroparticle Physics ENTApP
ILIAS/N6 under contract number RII3-CT-2004-506222 for financial
support. The use of the ``Ciclope'' cluster of the IFT-UAM/CSIC is also acknowledged.


\end{document}